\documentclass[twocolumn]{aastex63}

\usepackage{natbib}
\usepackage{amsmath}
\usepackage{multirow}
\usepackage[normalem]{ulem}
\usepackage{xcolor}
\usepackage{xspace}
\usepackage{enumitem}   
\usepackage{graphicx}
\usepackage{wrapfig}
\usepackage{lineno} 
\usepackage[super]{nth}

\newcommand{\Rearth}{$R_\oplus$\xspace}

\newcommand{\Msun}{$M_\odot$\xspace}
\newcommand{\wotan}{\texttt{W\={o}tan}\xspace}
\newcommand{\epos}{\texttt{epos}\xspace}
\newcommand{\eleanor}{\texttt{eleanor}\xspace}
\newcommand{\pterodactyls}{\texttt{pterodactyls}\xspace}
\newcommand{\tls}{\texttt{TLS}\xspace}

\newcommand{\gaia}{\texttt{Gaia}\xspace}
\newcommand{\kepler}{\emph{Kepler}\xspace}

\newcommand{\ktwo}{\emph{K2}\xspace}

\newcommand{\triceratops}{\texttt{triceratops}\xspace}
\newcommand{\exotic}{\texttt{EXOTIC}\xspace}
\newcommand{\rprs}{$\frac{R_\text{p}}{R_\star}$\xspace}
\newcommand{\muta}{$\mu$~Tau\xspace}
\newcommand{\PSUAA}{Department of Astronomy and Astrophysics, Penn State University, 525 Davey Laboratory, 251 Pollock Road, University Park, PA, 16802, USA}
\newcommand{\PSUCEHW}{Center for Exoplanets and Habitable Worlds, Penn State University, 525 Davey Laboratory, 251 Pollock Road, University Park, PA, 16802, USA}

\definecolor{tropicalrainforest}{rgb}{0.0, 0.46, 0.37}
\definecolor{plum}{HTML}{88498f}

\shorttitle{Young Cluster Occurrence}
\shortauthors{Fernandes, R.B. et al. 2025a}

\begin{document}
% \linenumbers
\title{Signatures of Atmospheric Mass Loss and Planet Migration in the Time Evolution of Short-Period Transiting Exoplanets}

\correspondingauthor{Rachel B. Fernandes}
\email{rbf5378@psu.edu}

\author[0000-0002-3853-7327]{Rachel B. Fernandes}
\altaffiliation{President's Postdoctoral Fellow}
\affil{\PSUAA}
\affil{\PSUCEHW}

\author[0000-0003-4500-8850]{Galen J. Bergsten}
\affil{Lunar and Planetary Laboratory, The University of Arizona, Tucson, AZ 85721, USA}

\author[0000-0002-1078-9493]{Gijs D. Mulders}
% \affil{Facultad de Ingenier\'ia y Ciencias, Universidad Adolfo Ib\'a\~nez, Av.\ Diagonal las Torres 2640, Pe\~nalol\'en, Santiago, Chile}
% \affil{Millennium Institute for Astrophysics, Chile}
\affil{Instituto de Astrof\'isica, Pontificia Universidad Cat\'olica de Chile, Av. Vicu\~na Mackenna 4860, 7820436 Macul, Santiago, Chile}

\author[0000-0001-7962-1683]{Ilaria Pascucci}
\affil{Lunar and Planetary Laboratory, The University of Arizona, Tucson, AZ 85721, USA}

\author[0000-0003-3702-0382]{Kevin K. Hardegree-Ullman}
\affil{Steward Observatory, The University of Arizona, Tucson, AZ 85721, USA}

\author[0000-0002-8965-3969]{Steven Giacalone}
\altaffiliation{NSF Astronomy and Astrophysics Postdoctoral Fellow}
\affil{Department of Astronomy, California Institute of Technology, Pasadena, CA 91125, USA}

\author[0000-0002-8035-4778]{Jessie L.\ Christiansen}
\affiliation{NASA Exoplanet Science Institute, IPAC, MS 100-22, Caltech, 1200 E.\ California Blvd, Pasadena, CA 91125}

\author[0000-0001-7615-6798]{James G. Rogers}
% \affil{Department of Earth, Planetary, and Space Sciences, The University of California, Los Angeles, 595 Charles E. Young Drive East, Los Angeles, CA 90095, USA}
\affil{Institute of Astronomy, University of Cambridge, Madingley Road, Cambridge CB3 0HA, United Kingdom}

\author[0000-0002-2006-7769]{Akash Gupta}
\altaffiliation{51 Pegasi b Fellow}
\altaffiliation{Future Faculty in Physical Sciences Fellow}
\altaffiliation{Harry H. Hess Postdoctoral Fellow}
\affil{Department of Astrophysical Sciences, Princeton University, Princeton, NJ 08544, USA}
\affil{Department of Geosciences, Princeton University, Princeton, NJ 08544, USA}

\author[0000-0001-9677-1296]{Rebekah I. Dawson}
\affil{\PSUAA}
\affil{\PSUCEHW}

\author[0000-0003-3071-8358]{Tommi T. Koskinen}
\affil{Lunar and Planetary Laboratory, The University of Arizona, Tucson, AZ 85721, USA}

\author[0000-0001-8153-639X]{Kiersten M. Boley}
\altaffiliation{NASA Sagan Fellow}
\affiliation{ Earth and Planets Laboratory, Carnegie Institution for Science, 5241 Broad Branch Road, NW, Washington, DC 20015, USA}

\author[0000-0002-2792-134X]{Jason L. Curtis}
\affil{Department of Astronomy, Columbia University, 550 West 120th Street, New York, NY 10027, USA}

\author[0000-0001-6476-0576]{Katia Cunha}
\affil{Steward Observatory, University of Arizona, 933 North Cherry Avenue, Tucson, AZ 85721-0065, USA}
\affil{Observat\'{o}rio Nacional/MCTIC, R. Gen. Jos\'{e} Cristino, 77, 20921-400, Rio de Janeiro, Brazil}

\author[0000-0003-2008-1488]{Eric E. Mamajek}
\affil{Jet Propulsion Laboratory, California Institute of Technology, 4800 Oak Grove Drive, Pasadena, CA 91109, USA}
\affil{Department of Physics and Astronomy, University of Rochester, Rochester, NY 14627-0171, USA}

\author[0000-0002-6650-3829]{Sabina Sagynbayeva}
\affil{Department of Physics and Astronomy, Stony Brook University, Stony Brook, NY 11794 USA}

\author[0000-0002-6673-8206]{Sakhee S. Bhure}
\affil{Centre for Astrophysics, University of Southern Queensland, Toowoomba, QLD, 4350, Australia}

\author[0000-0002-5741-3047]{David R. Ciardi}
\affil{NASA Exoplanet Science Institute, IPAC, California Institute of Technology, Pasadena, CA 91125 USA}

\author[0000-0002-1480-9041]{Preethi R. Karpoor}
\affil{Department of Astronomy \& Astrophysics, UC San Diego, La Jolla, CA, USA}

\author[0000-0002-5785-9073]{Kyle A. Pearson}
\affil{Jet Propulsion Laboratory, California Institute of Technology, 4800 Oak Grove Drive, Pasadena, CA 91109, USA}

\author[0000-0003-1848-2063]{Jon K. Zink}
\altaffiliation{NASA Hubble Fellow}
\affiliation{Department of Astronomy, Caltech, 1200 E. California Blvd, Pasadena, CA 91125}

\author[0000-0002-2012-7215]{Gregory A. Feiden}
\affil{Department of Physics and Astronomy, University of North Georgia, Dahlonega, GA 30597 USA}

% \author[0000-0001-6545-639X]{Eric B.\ Ford}
% \affil{\PSUAA}
% \affil{\PSUCEHW}
% \affil{\PSUICDS}
% \affil{\PSUCASt}

\begin{abstract}
Comparative studies of young and old exoplanet populations offer a glimpse into how planets may form and evolve with time. We present an occurrence rate study of short-period ($<$12\,days) planets between 1.8--10\,\Rearth around 1374 FGK stars in nearby (200\,pc) young clusters ($<$1\,Gyr), utilizing data from the Transiting Exoplanet Survey Satellite (TESS) mission. These planets represent a population closer to their primordial state. We find that the occurrence rate of young planets is higher ($64^{+32}_{-22}$\%) compared to the Gyr-old population observed by \kepler ($7.98^{+0.37}_{-0.35}$\%). Dividing our sample into bins of young (10--100\,Myr) and intermediate (100\,Myr--1\,Gyr) ages, we also find that the occurrence distribution in orbital period remains unchanged while the distribution in planet radius changes with time. Specifically, the radius distribution steepens with age, indicative of a larger planet population shrinking due to the atmospheric thermal cooling and mass loss. We also find evidence for an increase (1.9$\sigma$) in occurrence after 100\,Myr, possibly due to tidal migration driving planets inside of 12 days. While evidence suggests post-disk migration and atmospheric mass loss shape the population of short-period planets, more detections of young planets are needed to improve statistical comparisons with older planets. Detecting long-period young planets and planets $<$1.8\,\Rearth will help us understand these processes better. Additionally, studying young planetary atmospheres provides insights into planet formation and the efficiency of atmospheric mass loss mechanisms on the evolution of planetary systems.
\end{abstract}

%Additionally, we find that an increased occurrence rate of young planets in the hot Neptune desert when compared to \kepler, suggesting that planets that initially formed in this region may shrink below 1.8\,\Rearth due to atmospheric loss.

\section{Introduction} \label{sec:intro}
Prominent features in the the Gyr-old exoplanet population that we observe today are largely sculpted by the formation, migration, and evolutionary processes that take place in the early stages of planetary system formation. Two such examples are the hot Neptune desert \citep{Beauge2013}, and the radius valley \citep{fulton2017california, van2018asteroseismic, martinez2019spectroscopic, hardegree2020scaling}. While the hot Neptune desert is most likely sculpted by atmospheric mass loss processes such as XUV photoevaporation \citep{owen2013kepler,owen2017evaporation}, Roche lobe overflow \citep{Koskinen2022}, and/or core-powered mass loss \citep{ginzburg2016super,gupta2019sculpting,gupta2020corecool}, the origin of the radius valley is still hotly debated. It could be a product of (A) evolution in that it is created by atmospheric mass loss processes such as XUV photoevaporation and/or core-powered mass loss; (B) migration, which states that Neptune-sized planets may form farther from their host star, beyond the snow line where volatiles can condense into ice, but then migrate inward due to gravitational interactions with other planets or with the protoplanetary disk \citep{Bourrier2023,Venturini2020,Izidoro2022}; or (C) planet formation and envelope capture \citep{LeeConnors2021, perezbecker2013, izidoro2015}. One way to distinguish between these scenarios is to study young planets as they provide a sample much closer in time to when planets formed. By understanding what the planet population looks like at different ages, we can place constraints on the processes that dominate their formation, migration, and evolution.

%Young planets can also provide insights into the formation and evolution of our own solar system by comparing their properties with those of solar system planets. 

With the advent of wide-field, high-precision transit surveys such as NASA's \kepler \citep{Borucki2010}, \ktwo \citep{howell2014k2}, and TESS \citep{ricker2014transiting} missions, the study of young stars and their planets has gained significant momentum. Additionally, the Atacama Large Millimeter Array (ALMA) for high-resolution imaging of protostars and protoplanetary disks, along with direct imaging instruments on large optical and near-infrared telescopes, has made this research even more timely. The \ktwo mission, a follow-up to the highly successful \kepler mission, focused on the discovery of a variety of exoplanets in the ecliptic plane. While the mission's primary objective was not specifically to study young exoplanets, \ktwo made several notable detections of young exoplanets while surveying the Pleiades, Taurus, Hyades, Praesepe and Upper Scorpius clusters (e.g., \citealt{mann2016zodiacal_a, mann2016zodiacal_b, mann2017zodiacal, ciardi2018, rizzuto2017zodiacal, rizzuto2018zodiacal, gaidos2017zodiacal, vanderburg2018zodiacal}). Thorough investigations have already started to uncover the ways in which young planets differ from their Gyr-old counterparts. For example, \cite{Christiansen2023} studied the intermediate age (600--800\,Myr) clusters Praesepe and Hyades in which planets had previously been found using data from the \ktwo mission. For GKM host stars, they found a significantly elevated occurrence rate of hot ($<$10\,day), 1.8--4\,\Rearth sub-Neptunes (78--109\%) compared to the same parameter space for \kepler (16\%), which targeted older field stars. This substantial difference strongly indicates that such planets are far more abundant at younger stellar ages.

However, it was not until the TESS mission that we were able to sample a wide variety of young planets across different ages. During its two-year primary mission (PM), TESS observed 26 sectors of the sky for 27\,days each, covering about 85\% of the sky. This was followed by a three-year extended mission (EM1). In the first year of EM1, TESS revisited the four clusters observed by \ktwo, then spent the remaining two years observing the southern and northern hemispheres. Throughout PM and EM1, TESS obtained Full Frame Images (FFI) in 30-min and 10-min cadence modes, respectively, providing a relatively untapped reservoir of young planetary systems. In fact, the discovery of several young planets using TESS has revealed that the young planets are larger in size, when compared to those in the Gyr-old population (e.g., \citealt{newton2019tess,newton2021tess, rizzuto2020tess,mann2020tess,nardiello2020psf,bouma2020cluster}). However, this could be an observational bias where the high rotational variability in the light curves of younger stars makes it difficult to recover smaller planets. Therefore, in order to truly understand this population of young planets, we need to study them in an unbiased, demographic manner.

In \citet{fernandes2022} and \cite{fernandes_khu_2023}, we computed intrinsic occurrence rates for five clusters with known transiting planets: Tucana-Horologium Association, IC~2602, Upper Centaurus Lupus, Ursa Major, and Pisces--Eridanus. We found an intrinsic planet occurrence rate of 90$\pm$37\% for sub-Neptunes and Neptunes (1.8--6\,\Rearth) orbiting within 12.5\,days of young FGK stars, consistent with that of \cite{Christiansen2023}, and significantly higher than \kepler's Gyr-old rate of 7.98$\pm$0.33\% in the same bin. However, it is important to note that the intrinsic planet occurrence rates in both \cite{fernandes_khu_2023} and \cite{Christiansen2023} are biased given that they studied only clusters with confirmed/candidate planets. More recently, \cite{Vach2024} independently analysed an unbiased volume-limited sample of 7219 young stars in co-moving groups younger than 200\,Myr that were observed in TESS's first four years, searching for 2--8\,\Rearth, short-period (1.6--20\,days) transiting planets. They computed an occurrence rate of $22^{+8.6}_{-6.8}$\% for sub-Neptunes (2--4\,\Rearth), and $13^{+3.9}_{-4.9}$\% for super-Neptunes (4--8\,\Rearth), respectively, from the TESS FFI data, suggesting that short-period young planets with these radii are more abundant than their Gyr-old counterparts. It is worth highlighting that \cite{Vach2024} is limited to $<$200\,Myr, and therefore we do not have an unbiased understanding of clusters at intermediate ages (up to 1\,Gyr).

In this work, we bridge the gap between the young (10--100\,Myr) and intermediate (100\,Myr--1\,Gyr) aged planetary systems. By building on the methodology established in \cite{fernandes2022} and \cite{fernandes_khu_2023}, we compute the intrinsic occurrence of short-period planets in young clusters (Section~\ref{sec:methodology}). In Section~\ref{sec:recovery}, we search for transiting young confirmed planets and planet candidates in our sample of 31 young clusters and moving groups within 200\,pc. Here, we also discuss how well we recover known young transiting confirmed planets and planet candidate signals in TESS EM1 FFIs using our pipeline, \pterodactyls. We then combine these recovered planets with our homogeneously-derived catalog of updated stellar parameters \citep{fernandes_khu_2023} to improve upon our previous estimates of their intrinsic occurrence rate around Sun-like (FGK) stars (Section~\ref{sec:occurrence}), and place our results into context by comparing them to planet formation, migration and evolution theories in Section~\ref{sec:results}. Finally, in Section~\ref{sec:summary}, we discuss how performing a quantitative comparison with \kepler's Gyr-old FGK short period population can establish how the period and radius distribution of short-period transiting exoplanets has evolved over time, and therefore provide observational constraints on migration, and  atmospheric mass loss processes that affect the evolution of planetary systems with time.

%%%%%%%%%%%%%%%%%%%%%%%%%%%%%%%%%%%%%%%%%%%%%%%%%%%%%%%
%%%%%%%%%%%%%%%%%%% Methodology %%%%%%%%%%%%%%%%%%%%%%%
%%%%%%%%%%%%%%%%%%%%%%%%%%%%%%%%%%%%%%%%%%%%%%%%%%%%%%%
\section{Methodology}\label{sec:methodology}

\begin{figure*}[!htb]
    \centering
    \includegraphics[width=\linewidth]{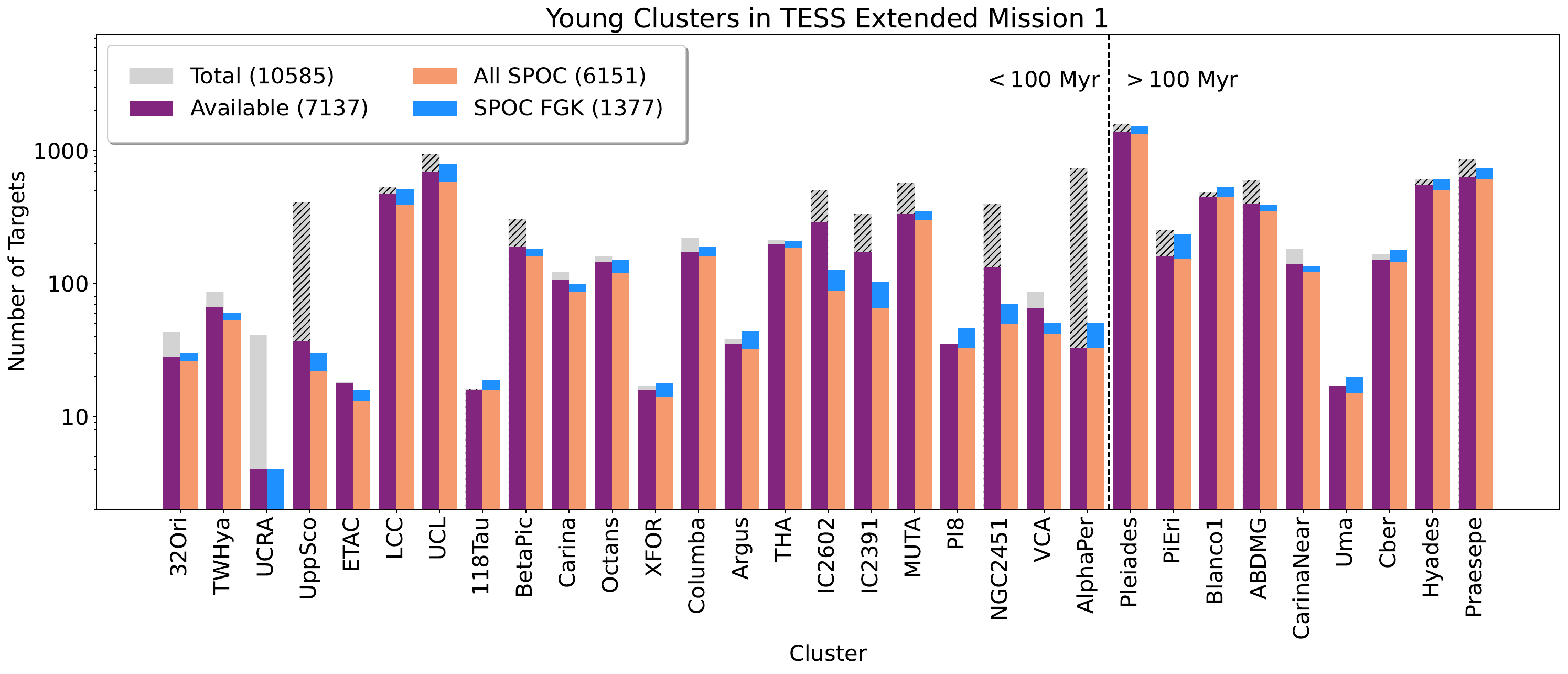}
    \caption{Bar chart showing the distribution of targets in young clusters for TESS EM1, highlighting the distinction between clusters younger than 100\,Myr and those older than 100\,Myr. The bars represent the total number of members in the clusters (grey dashed lines), targets observed during TESS EM1 (orange), targets that have SPOC light curves available (purple), and FGK targets for which SPOC light curves are available (blue).}
\label{fig:sample_histogram}
\end{figure*}

In \citet{fernandes2022}, we created \pterodactyls{} $-$ a specialized tool curated to search and vet transiting planetary signals in the highly variable TESS light curves of young stars. We also computed preliminary occurrence rates of a test sample of five clusters with known transiting planets, at the time: Tucana-Horologium Association, IC~2602, Upper Centaurus Lupus, Ursa Major and Pisces--Eridanus. For this biased sample, we conducted injection-recovery tests to characterize our detection efficiency, and computed an intrinsic planet occurrence rate of 49$\pm$20\% for sub-Neptunes and Neptunes (1.8--6\,\Rearth) with orbital periods less than 12.5\,days, which is higher than \kepler's Gyr-old occurrence rates of 7.98$\pm$0.33\%. At the time, since we did not have a uniform set of stellar parameters, we conducted our injection-recovery tests and occurrence rate calculations in orbital period-\rprs space. Since exoplanet parameters depend on the properties of their host stars, it is important to properly characterize the host stars since any systematic biases in the derivation of host star parameters can negatively impact the derivation of planetary parameters. To address this issue, in \citet{fernandes_khu_2023}, we created a uniform catalog of photometrically-derived stellar effective temperatures, luminosities, radii, and masses for 4,876 of the young stars in our sample. With these stellar radii, we updated our injection-recovery tests to orbital period-planet radius space. This also allowed us to focus solely on the young FGK stars in our sample for a better comparison to \kepler’s FGK occurrence rates. As part of this refinement, we excluded M dwarfs, which tend to have lower detection efficiencies due to their faintness and higher activity levels. This resulted in a 1.5x increase in detection efficiencies for short-period sub-Neptunes/Neptunes (1.8--6\,\Rearth) compared to our analysis in \cite{fernandes2022}.

% When stellar parameters were taken into account, we saw a $1.5\times$ increase in the detection efficiencies for short-period sub-Neptunes/Neptunes (1.8--6\,\Rearth) implying that, for our sample of young FGK stars, better characterization of host star properties can lead to the recovery of small transiting planets. 

For this work, we expanded our search to 31 young clusters, moving groups and associations within 200\,pc. Here, we use data from the TESS EM1 FFIs (10-min cadence) since most of the stars in our sample do not have 2-min cadence data. A significant advantage of EM1 data is its unique coverage of the \ktwo campaigns, incorporating four young clusters (Upper Scorpius, Pleiades, Hyades, and Praesepe), facilitating the recovery of planets initially detected by \ktwo using TESS. 

\subsection{Sample of Young Stellar Clusters} \label{sec:sample}
We assembled a sample of nearby moving groups and young clusters from the BANYAN $\Sigma$ \citep{gagne2018banyan} and \gaia DR2 open cluster member lists \citep{babusiaux2018gaia}. Additionally, we included the Argus \citep{Zuckerman2019}, \muta \citep{gagne2020}, and Pisces Eridanus \citep{Curtis2019} groups. We restricted the distance to approximately $\sim$200\,pc, as Sun-like stars (particularly fainter K dwarfs) beyond this range are too faint for TESS planet detection, i.e., Neptunes can no longer be detected above the noise in the TESS light curves. Stars in clusters younger than 10\,Myr were excluded due to potential disk retention \citep[e.g.,][]{ErcolanoPascucci2017} and their intricate, variable light curves \citep[e.g.,][]{Cody2014}. Clusters older than $\sim$1\,Gyr were also omitted as most of the atmospheric mass loss is expected to have occurred within this timescale \citep{Owen2024,Rogers2024}. With these criteria, we initially identified 10,585 young stars across 31 young clusters and moving groups.

\subsection{\pterodactyls}
In our previous works \citep{fernandes2022, fernandes_khu_2023}, we employed PSF light curves extracted using \eleanor \citep{feinstein2019eleanor}. For this study, we opted for the TESS SPOC FFI PDCSAP light curves from \citet{Caldwell2020} instead due to their lower RMS residual scatter compared to those from \eleanor, enhancing our ability to detect small planets in these inherently variable light curves (see Appendix~\ref{app:lc_comp} for more details). Of these, 7137 stars were observed during TESS EM1 (see Figure~\ref{fig:sample_histogram}), and 6151 have available SPOC light curves. 5457 of the targets with SPOC light curves are usable i.e., did not have significant instrumental effects, and were searched for planet signals. After removing OBA stars, white dwarfs, late-type M dwarfs, and probable binaries as identified by \citet{fernandes_khu_2023}, we were left with 2777 young stars. Because we are interested in studying the population of short-period planets around Sun-like stars (and to enable direct comparisons with \kepler's Sun-like sample), we only focus on 1374 FGK-type stars for which we have homogeneously derived stellar parameters \citep{fernandes_khu_2023}.

While the TESS SPOC pipeline effectively mitigates most TESS systematics, certain instrumental effects remain in these light curves. These include mid-sector flux drops from data downlink and offsets due to errors in the uploaded Guidestar tables (see TESS’ Data Release Notes for details\footnote{\url{https://archive.stsci.edu/tess/tess_drn.html}}). Given that these effects could create spikes in the light curve, impeding \pterodactyls' ability to effectively detrend and detect planetary signals in our search, we refined the light curves by masking the mid-sector flux drops and adjusting offsets. We also masked several cadences that have an elevated number of transit-like detections, indicating issues with the light curves in those segments (see Figures~\ref{fig:sem_y3} and \ref{fig:sem_y4} in Appendix~\ref{app:sem}). Following the approach in \cite{fernandes2022}, we utilized a test similar to the ``Skye” metric by \cite{thompson2018} for \kepler and adapted for \ktwo data by \cite{zink2020scaling} to evaluate transit-like signals at each cadence. If the signal count exceeded 3$\sigma$, we masked those problematic cadences prior to rerunning the search.

Despite transitioning from PM's 30-min cadence \eleanor PSF light curves to EM1's 10-min cadence TESS SPOC light curves from FFIs, we found that the majority of the pipeline is not changed. In particular, the effectiveness of our detrending method i.e., a penalized spline optimized using rotation rate in \wotan \citep{hippke2019wotan}, planet search using \texttt{transitleastsquares} (\tls; SNR=SDE$\geq$7), and vetting and validation tests \citep[\texttt{EDI-Vetter Unplugged;}][]{zink2019edivetter} and \triceratops \citep{giacalone2020vetting} remain unaffected. As such, no changes were made to them.

\subsection{Accounting for Flux Contamination}
The radii of transiting exoplanets cannot be directly measured. Instead, their value is estimated from the transit depth, which is approximately equal to the ratio between the square of the planet's radius and the host star's radius. However, in the case of crowded fields (typical in young cluster environments), it is highly probable that a given transit is diluted due to the light from a bound companion or a line-of-sight, unbound star. Such a dilution of the transit depth would further lead to an underestimated, and hence inaccurate, measurement of the planetary radii which in turn would affect how a planet is characterized. In fact, the dilution of the transit signal by unresolved stars in the \kepler field led to an underestimation of the planet radius by a factor of $\sim$1.5 on average (see e.g., \citealt{ciardi2015understanding,teske2018effects}).

Stellar companions, both bound and line-of-sight, decrease the ability of the pipelines to detect transits and to properly measure planetary radii. As demonstrated with the \kepler survey and follow-up high-resolution imagery of exoplanet candidate host stars, a proper treatment of photometric blending decreases the number of planets $\leq$2\,\Rearth{} by up to $\sim$20\% and increases the number of planets $>$6\,\Rearth{} by up to $\sim$68\% \citep{furlan2017kepler}. Since stars in young clusters are in close proximity to each other, there is an increased chance of photometric blending. This is even more important for TESS observations where the pixel scale is 21$''$, and in the environment of stellar clusters where much of the blending is likely from other stars in the cluster that are not bound to the host star. This is of particular importance to our work since in order to understand planetary evolution, we need a good understanding of the planet radius distribution. 

However, multiplicity rates in young clusters have not been well determined in the sense that they have only been calculated for a handful of clusters (see, for e.g., \citealt{duchene2013stellar, Offner2023}). Given that, we settle for a first-order approximation of the flux contamination in our clusters by using \triceratops's flux contamination mode \citep{giacalone2020vetting} to calculate the amount of flux each star is contributing to a fixed aperture by taking into account nearby, background and line-of-sight stars from \gaia DR3.

\section{Planet Search and Recovery}\label{sec:recovery}
\begin{figure*}[!htpb]
\centering
\begin{minipage}{0.48\linewidth}
  \centering
  \includegraphics[width=\linewidth]{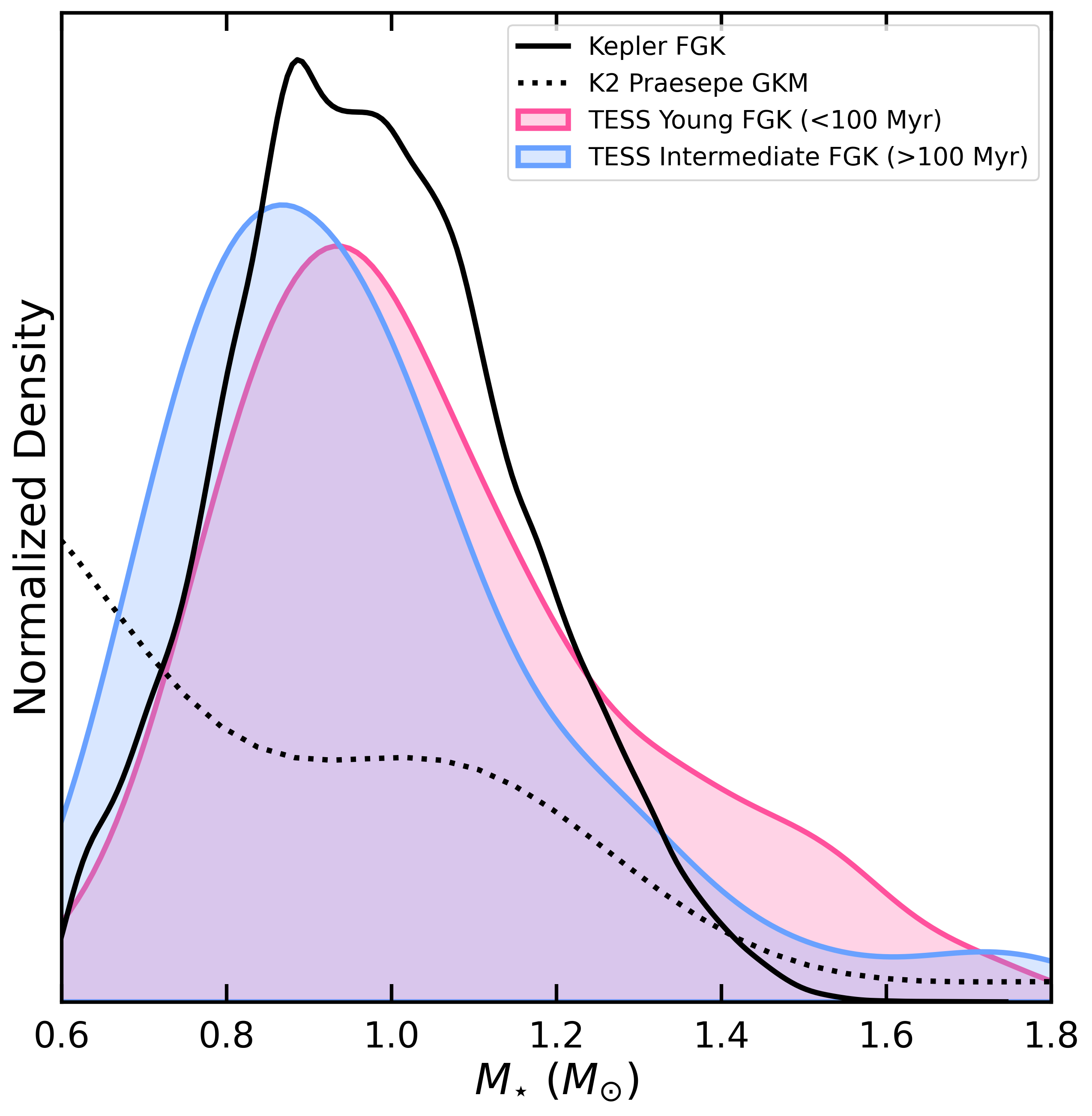}
\end{minipage}
\hfill
\begin{minipage}{0.48\linewidth}
  \centering
  \includegraphics[width=\linewidth]{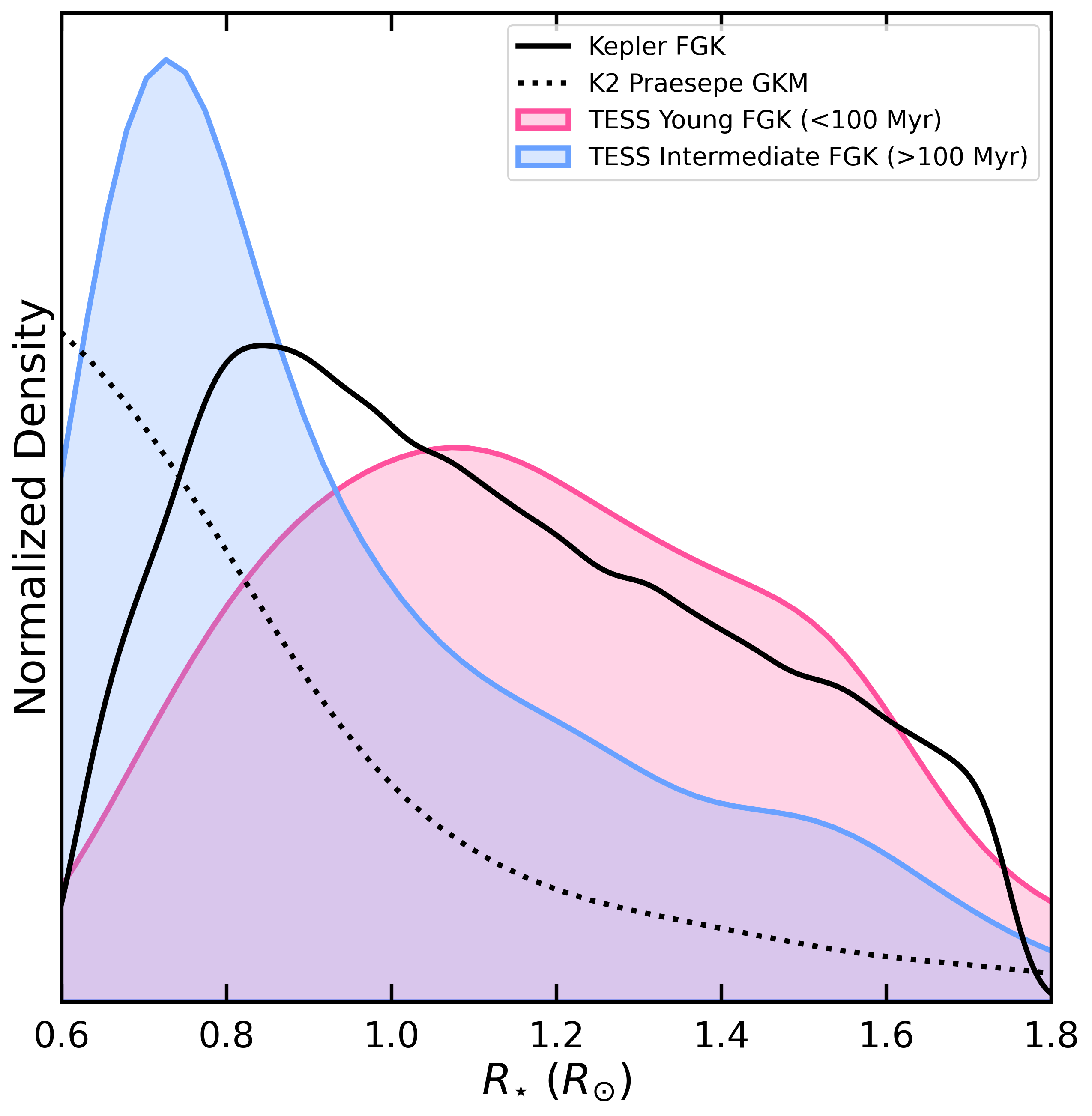}
\end{minipage}
\vspace{0.5em} % Adjust spacing between rows
\begin{minipage}{0.48\linewidth}
  \centering
  \includegraphics[width=\linewidth]{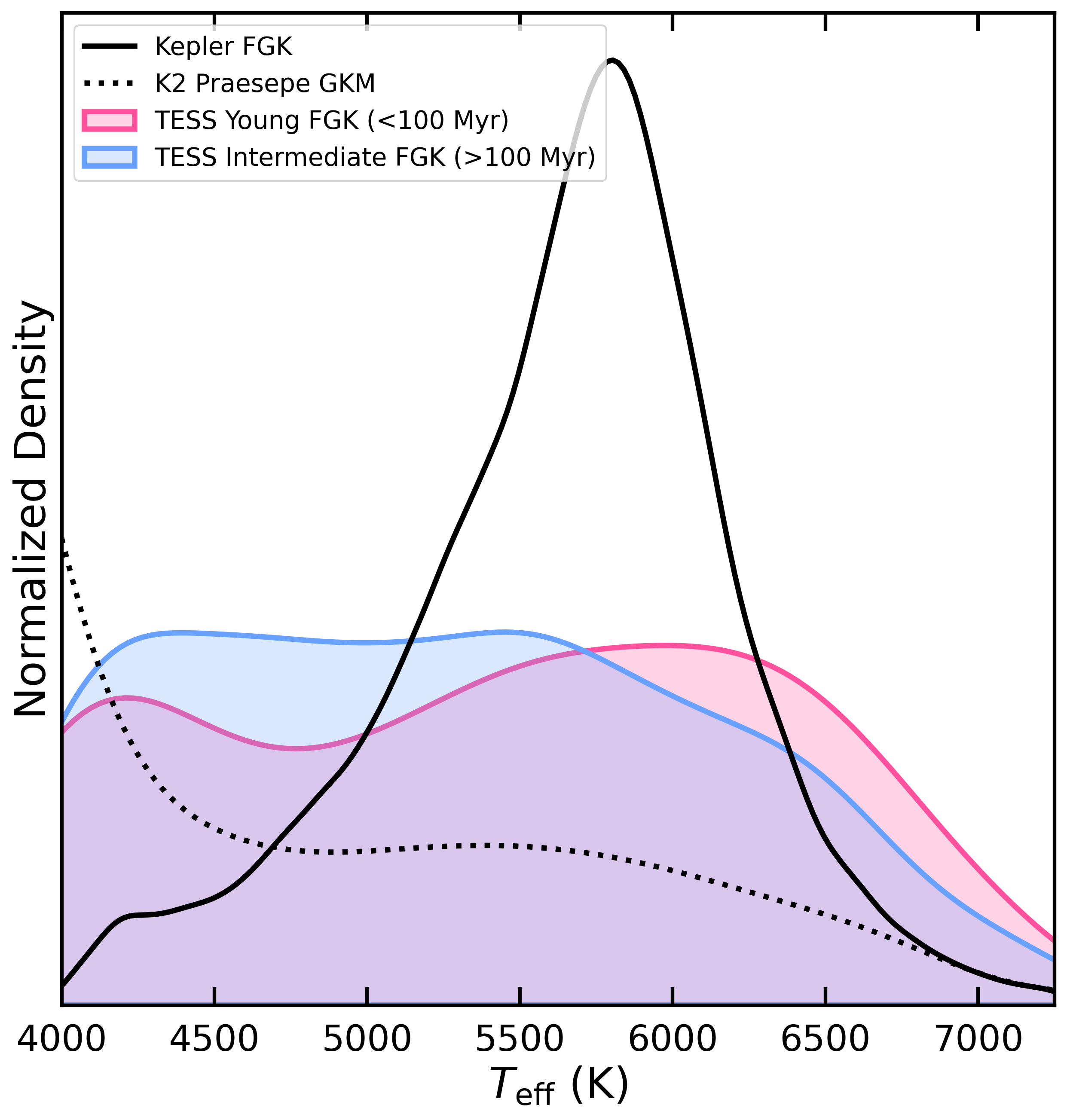}
\end{minipage}
\hfill
\begin{minipage}{0.48\linewidth}
  \centering
  \includegraphics[width=\linewidth]{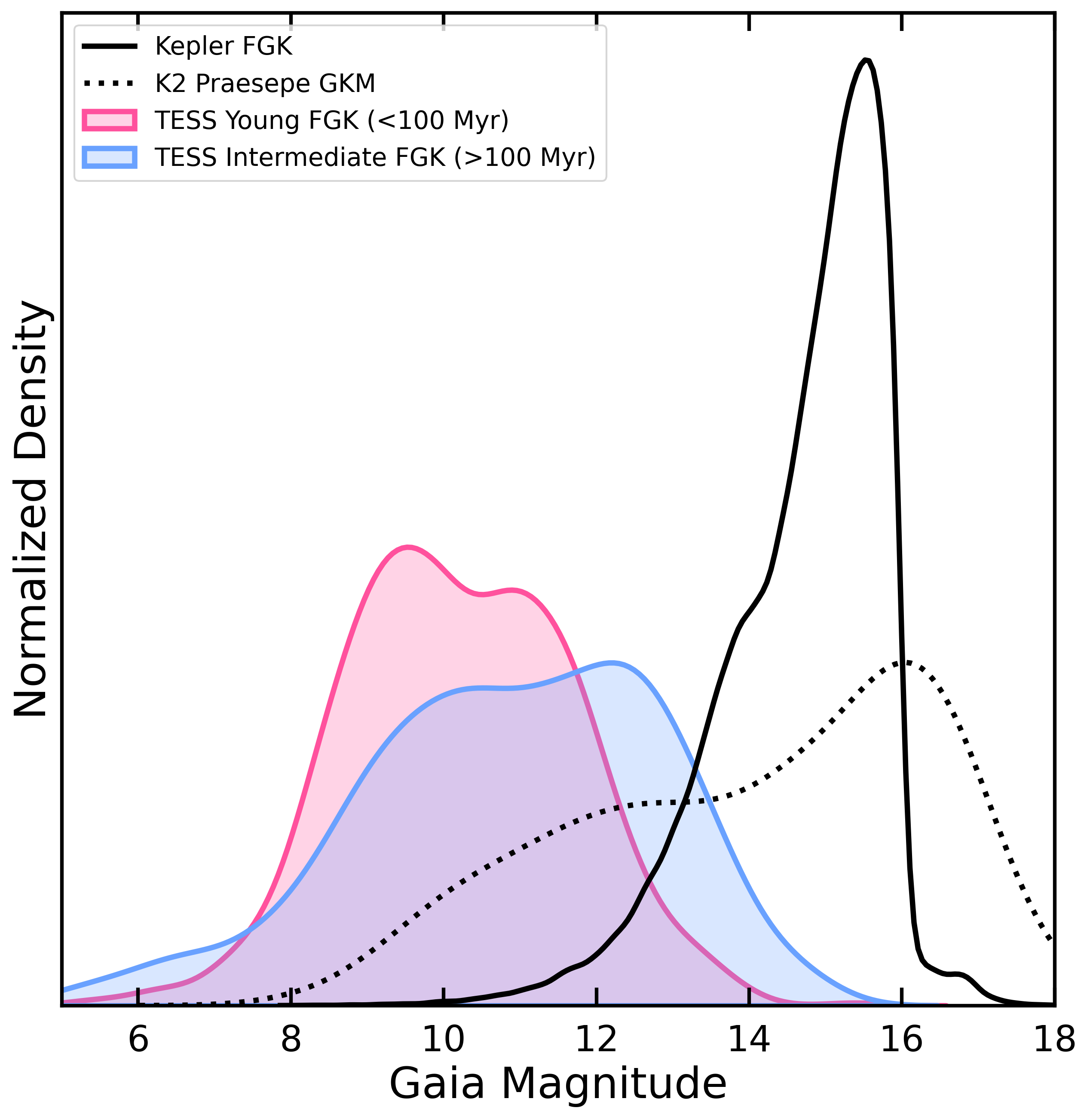}
\end{minipage}
\caption{Normalized density distributions of stellar properties such as stellar mass (top left), radius (top right), effective temperature (bottom left), and Gaia magnitude (bottom right) for host stars from \kepler (black solid lines), \cite{Christiansen2023}'s \ktwo Praesepe (black dashed lines), and TESS (pink and blue shaded regions for young and intermediate-age stars, respectively).}
\label{fig:sample_comp}
\end{figure*}
In this study, we only focus on the TESS EM1 FFI (10-min cadence) data to mitigate any effects that different cadences can have on our planet recovery and demographic analyses. In our young stellar sample, there are 32 confirmed planets and planet candidates in 23 systems discovered using both the \ktwo and TESS missions, out of which 28 were observed in TESS EM1 (see  Table~\ref{table:known_planets}). We processed all of the 5457 young stellar light curves extracted from TESS EM1 FFIs through \pterodactyls, successfully recovering 16 planets. All our recovered planets passed a Threshold Crossing Event (TCE) cut with a signal-to-noise ratio (SNR) of 7, and a signal detection efficiency (SDE) of 7. Additionally, all recovered planets passed the following statistical vetting tests in \pterodactyls: at least two transits with data, orbital period dissimilar to the stellar rotation rate, consistency in individual transit depths, observed transit duration consistent with expected transit duration, and \texttt{EDI-Vetter}'s Individual Transit Test and Secondary Eclipse Test \citep{zink2019edivetter}. Finally, the phase-folded light curves of all recovered planets were validated using \triceratops \citep{giacalone2020vetting}. When determining the probability of a transiting signal being an astrophysical false positive or a transiting planet, \triceratops (by default) uses planet occurrence rate priors based on the Gyr-old exoplanet population (e.g., \citealt{howard2012planet, dongzhu2013, petigura2013prevalence, dressing2015occurrence, mulders2015stellar, Mulders2018}). However, the planet population around young stars is likely different from that around Gyr-old stars; for example, young planets are typically found to be larger in size. This difference can bias the probability calculations made by \triceratops. Therefore, we also calculated the astrophysical false positive probability using non-informative (flat) priors recently implemented in \triceratops.

Out of the 12 planets we were unable to recover, 9 were \ktwo discoveries that TESS was unable to recover: \ktwo-136\,b, c, and d, HD 283869\,b, \ktwo-102\,b, \ktwo-103\,b, \ktwo-104\,b, and \ktwo-264\,b and c. This is because \ktwo had better precision and a longer baseline than TESS, allowing it to detect long-period planets around fainter stars that TESS lacks the sensitivity to find. Additionally, we could not recover 3 planets (TOI 1227\,b, HIP 67522\,c, and AU Mic\,c) that were either in multi-planet systems where only one planet was recoverable due to a combination of light curve quality, and/or long period against TESS's short baseline and small planet size.

In total, we recovered 16 confirmed planets and planet candidates, 14 of which are around Sun-like (FGK) stars, and are included in our final FGK sample of 1374 stars. However, two of these planets have orbital periods $>$12\,days, and one planet has a radius $<1.8$\,\Rearth. There are 4 planets in our young (10--100\,Myr) age bin, and 7 planets in our intermediate (100\,Myr--1\,Gyr) age bin, leaving us with 11 planets that we have used in our occurrence rate calculations below. A comparison of our young and intermediate age stellar sample compared to \kepler and \cite{Christiansen2023}'s \ktwo Praesepe sample is presented in Figure~\ref{fig:sample_comp}.

\begin{table*}[!htb]
    \centering
    \begin{tabular}{|c|c|c|c|c|c|c|c|}
    \hline
        Planet Recovered & TIC ID & Cluster (Age in Myr) & Radius$^{*}$ (\Rearth) & Period$^{*}$ (days) & FPP (Gyr-old) & FPP (flat)  \\ \hline
        % HD~109833\,b$^{a}$ & 360630575 & LCC (15$\pm$3) & 2.89$\pm$0.15 & 9.18$\pm$0.15 & 0.022$\pm$0.003 & 0.089$\pm$0.015\\      
        HIP~67522\,b$^{a}$ & 166527623 & UCL (16$\pm$2) & $9.72^{+0.48}_{-0.47}$ & 6.964$\pm$0.026 & 0.88$\pm$0.02 & 0.75$\pm$0.02  \\ 
        % TOI~4333\,b$^{c}$ & 144645881 & UCL (16$\pm$2) & 9.09$\pm$0.02 & 8.91$\pm$0.04 & 0.76$\pm$0.02 & 0.47$\pm$0.02\\ 
        AU~Mic\,b$^{b}$ & 441420236 & BetaPic (24$\pm$3) & 4.02$\pm$0.21 & 8.45$\pm$0.04 & 0.0043$\pm$0.0008 & 0.021$\pm$0.002  \\  
        DS Tuc A\,b$^{c}$ & 410214986 & THA (45$\pm$4) & 5.70$\pm$0.17 & 8.138$\pm$0.023 & 0.25$\pm$0.03 & 0.27$\pm$0.02  \\ 
        TIC~460950389\,b$^{d}$ & 460950389 & IC~2602 ($46^{+6}_{-5}$) & 3.8$\pm$0.2 & 2.862$\pm$0.009 & 0.0232$\pm$0.0006 & 0.075$\pm$0.002  \\ 
        TOI~837\,b$^{d}$ & 460205581 & IC~2602 ($46^{+6}_{-5}$) & $6.9^{+0.6}_{-0.4}$; $8.90^{+0.74}_{-0.71}$ & 8.325$\pm$0.016 & 0.11$\pm$0.06 & 0.03$\pm$0.01  \\ 
        TOI~5358\,b$^{e}$ & 46631742 & Pleiades (112$\pm$5) & 3.52$\pm$0.37 & 2.659$\pm$0.003 & 0.018$\pm$0.001 & 0.081$\pm$0.004  \\    
        TOI~451\,b$^{f}$ & 257605131 & PscEri ($\sim$120) & $1.94^{+0.15}_{-0.34}$ & 1.857$\pm$0.008 & 0.057$\pm$0.001 & 0.139$\pm$0.004  \\ 
        TOI~451\,c$^{f}$ & 257605131 & PscEri ($\sim$120) & 3.07$\pm$0.14 & 9.19$\pm$0.06 & 0.0330$\pm$0.0009 & 0.127$\pm$0.005  \\ 
        TOI~451\,d$^{f}$ & 257605131 & PscEri ($\sim$120) & 4.03$\pm$0.15 & 16.36$\pm$0.11 & 0.005$\pm$0.001 & 0.032$\pm$0.005  \\ 
        TOI~4399\,b$^{g}$ & 464646604 & ABDMG ($149^{+51}_{-19}$) & $3.00^{+0.32}_{-0.28}$ & 7.70$\pm$0.04 & 0.056$\pm$0.004 & 0.16$\pm$0.01  \\ 
        TOI~1726\,b$^{h}$ & 130181866 & UMa (414$\pm$23) & 2.15$\pm$0.10 & 7.11$\pm$0.01 & 0.011$\pm$0.002 & 0.011$\pm$0.006  \\ 
        TOI~1726\,c$^{i}$ & 130181866 & UMa (414$\pm$23) & 2.67$\pm$0.12 & 20.54$\pm$0.04 & 0.006$\pm$0.003 & 0.015$\pm$0.013  \\ 
        TOI~1726\,d$^{i}$ & 130181866 & UMa (414$\pm$23) & 1.084$\pm$0.043 & 4.209$\pm$0.005 & 0.044$\pm$0.002 & 0.049$\pm$0.010  \\
        K2-25\,b$^{j}$ & 434226736 & Hyades (750$\pm$100) & $3.43^{+0.95}_{-0.31}$ & 3.484552$\pm$0.000044 & 0.012$\pm$0.003 & 0.022$\pm$0.005   \\
        TOI~4364\,b$^{k}$ & 4070275 & Hyades (750$\pm$100) & $2.10^{+0.10}_{-0.08}$ & 5.424019$\pm$0.000011 & 0.061$\pm$0.005 & 0.056$\pm$0.003  \\
        K2-100\,b$^{l}$ & 307733361 & Praesepe (790$\pm$60) & $3.5^{+0.2}_{-0.2}$ & 1.673$\pm$0.002 & 0.015$\pm$0.002 & 0.084$\pm$0.008  \\ \hline

        Planet Not Recovered & TIC ID & Cluster (Age in Myr) & Radius$^{+}$ (\Rearth) & Period$^{+}$ (days) & FPP (Gyr-old) & FPP (flat)  \\ \hline
        % HD~109833\,c$^{a}$ & 360630575 & LCC (15$\pm$3) & 2.63$\pm$0.12 & 13.90014$\pm$0.00003 & X & X\\
        TOI~1227\,b$^{m}$ & 360156606 & LCC (15$\pm$3) & 9.5$\pm$0.5 & 27.36397$\pm$0.00011 & X & X\\         
        HIP~67522\,c$^{n}$ & 166527623 & UCL (16$\pm$2) & $8.01^{+0.75}_{-0.71}$ & $54^{+70}_{-0.24}$ & X & X \\          
        AU~Mic\,c$^{o}$ & 441420236 & BetaPic (24$\pm$3) & 3.51$\pm$0.16 & $\sim$18.86 & X & X  \\ 
        % K2~77\,b$^{n}$ & 435339847 & Pleiades (112$\pm$5) & 2.30$\pm$0.16 & 8.1998$\pm$0.0007 & X & X \\    
        K2-136\,b$^{p}$ & 18310799 & Hyades (750$\pm$100) & $0.99^{+0.06}_{-0.04}$ & $7.975292^{+0.000833}_{-0.000770}$ & X & X\\ 
        K2-136\,c$^{p}$ & 18310799 & Hyades (750$\pm$100) & $2.91^{+0.11}_{-0.10}$ & $17.307137^{+0.000252}_{-0.000284}$ & X & X \\  
        K2-136\,d$^{p}$ & 18310799 & Hyades (750$\pm$100) & $1.45^{+0.11}_{-0.08}$ & $25.575065^{+0.002418}_{-0.002357}$ & X & X  \\   
        HD~283869\,b$^{q}$ & 59873985 & Hyades (750$\pm$100) & 1.96$\pm$0.13 & $106^{+74}_{-25}$ & X & X  \\        
        K2-102\,b$^{l}$ & 175262071 & Praesepe (790$\pm$60) & $1.3^{+0.1}_{-0.1}$ & $9.915615^{+0.001209}_{-0.001195}$ & X & X\\    
        K2-103\,b$^{l}$ & 337632006 & Praesepe (790$\pm$60) & $2.2^{+0.2}_{-0.1}$ & $21.169619^{+0.001665}_{-0.001729}$ & X & X\\ 
        K2-104\,b$^{l}$ & 175194958 & Praesepe (790$\pm$60) & $1.9^{+0.2}_{-0.1}$ & $1.974190^{+0.000110}_{-0.000110}$ & X & X\\         
        K2~264\,b$^{r}$ & 184914317 & Praesepe (790$\pm$60) & $2.27^{+0.20}_{-0.16}$ & $5.839770^{+0.000063}_{-0.000061}$ & X & X \\ 
        K2~264\,c$^{r}$ & 184914317 & Praesepe (790$\pm$60) & $2.77^{+0.20}_{-0.18}$ & $19.663650^{+0.000303}_{-0.000306}$ & X & X \\  \hline

        Planet Not Observed & TIC ID & Cluster (Age in Myr) & Radius$^{+}$ (\Rearth) & Period$^{+}$ (days) & FPP (Gyr-old) & FPP (flat)  \\ \hline
        K2-33\,b$^{s}$ & 49040478 & UppSco (10$\pm$3) & $5.04^{+34}_{-37}$ & $5.424865^{+0.000035}_{-0.000031}$ & X & X \\        
        K2-95\,b$^{l}$ & 195193025 & Praesepe (790$\pm$60) & $3.7^{+0.2}_{-0.2}$ & $10.135091^{+0.000495}_{-0.000488}$ & X & X \\
        EPIC~211901114\,b$^{l}$ & 175291050 & Praesepe (790$\pm$60) & $9.6^{+5.3}_{-4.8}$ & $1.648932^{+0.000071}_{-0.000069}$ & X & X  \\       
        K2-101\,b$^{l}$ & 175265494 & Praesepe (790$\pm$60) & $2.0^{+0.1}_{-0.1}$ & $14.677286^{+0.000828}_{-0.000804}$ & X & X \\ \hline       
    \end{tabular}
    \caption{List of confirmed planets and planet candidates in our sample of young stellar clusters and moving groups. Age and cluster memberships are from \cite{gagne2018banyan} and \cite{babusiaux2018gaia}. $^{*}$Recovered radii and orbital periods consistent with published literature values; $^{+}$Published radii and orbital periods due to non-recovery. False Positive Probability (FPP) is derived using \triceratops assuming both Gyr-old and flat priors.
    References: (a) \cite{rizzuto2020tess}; (b) \cite{Plavchan2020}; (c) \cite{newton2019tess}; (d) \cite{nardiello2020psf}; (e) ExoFOP; (f) \cite{newton2021tess}; (g) \cite{zhou2022}; (h) \cite{Capistrant2024}; (i) \cite{mann2020tess}; (j) \cite{mann2016zodiacal_a}; (k) \cite{Distler2024}; (l) \cite{mann2017zodiacal}; (m) \cite{Mann2022}; (n) \cite{Barber2024};  (o) \cite{Cale2021}; (p) \cite{Mann2018}; (q) \cite{vanderburg2018zodiacal}; (r) \cite{rizzuto2018zodiacal}; (s) \cite{mann2016zodiacal_b}. }
    \label{table:known_planets}    
\end{table*}

\section{Intrinsic Occurrence of Young Planets}\label{sec:occurrence}
\begin{figure*}[!htb]
    \centering
    \includegraphics[width=\linewidth]{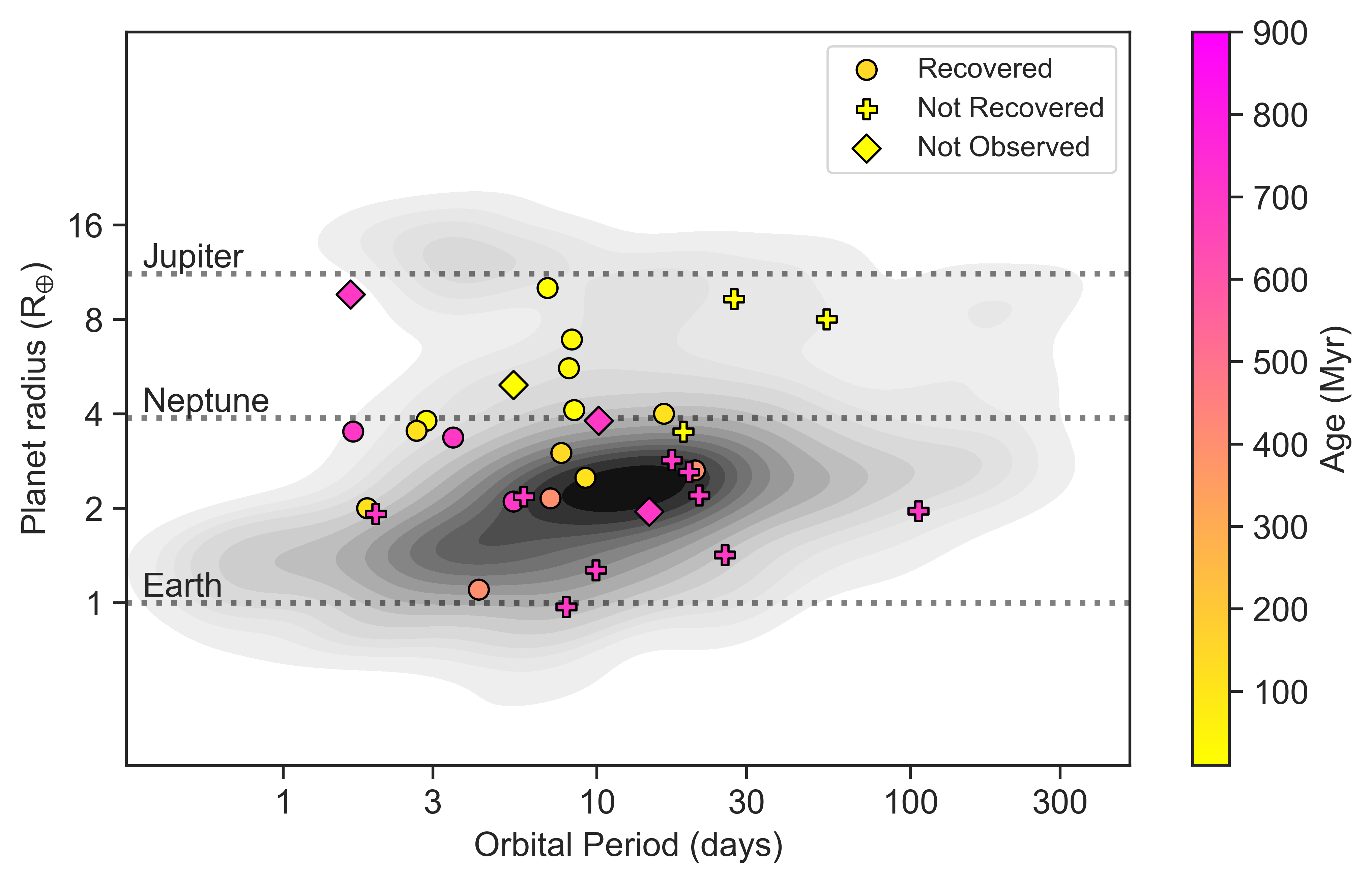}
    \caption{Comparison of the young transiting planet population in our sample (color-coded by age), and the Gyr-old transiting planet population (gray contours). Young gas-rich planets appear to be puffier, and ``shrink'' as they age, thereby populating gaps in the Gyr-old distribution, as indicated by our statistical analysis of this sample.}
\label{fig:young_old}
\end{figure*}
As can be seen in Figure~\ref{fig:young_old}, the unbiased radial distribution of short-period sub-Neptunes and Neptunes appears to be inflated at young ($<$100\,Myr) ages, and to be shrinking with time. The planet sample, however, is small and it is also possible that we are only able to recover larger planets in the light curves because younger stars are more variable. Therefore, we need to carefully calculate the intrinsic occurrence rates of transiting planets in young clusters and moving groups, and compare them to that of \kepler, in order to verify whether young planets indeed ``shrink" over time due to thermal evolution and atmospheric mass loss.

\subsection{Characterizing Survey Completeness}\label{subsec:completeness}

\begin{figure}[!htpb]
\minipage{\linewidth}
  \includegraphics[width=\linewidth]{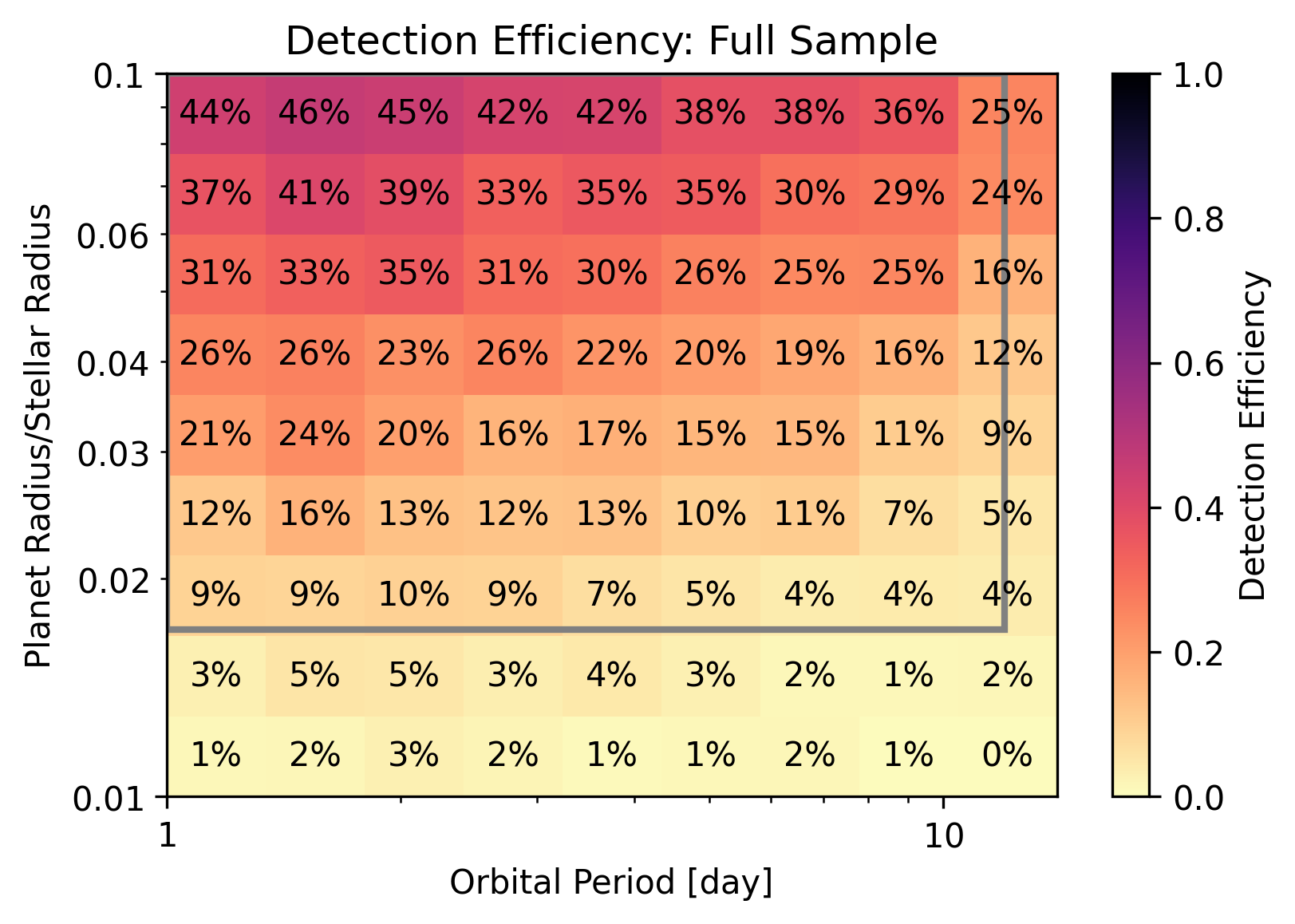}
\endminipage\hfill
\minipage{\linewidth}
  \includegraphics[width=\linewidth]{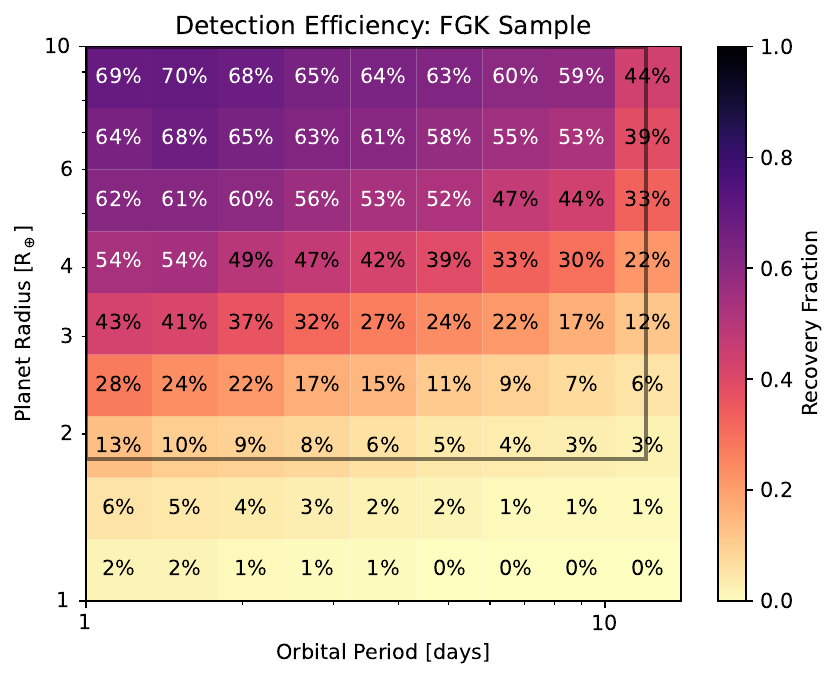}
\endminipage\hfill
\minipage{\linewidth}
  \includegraphics[width=\linewidth]{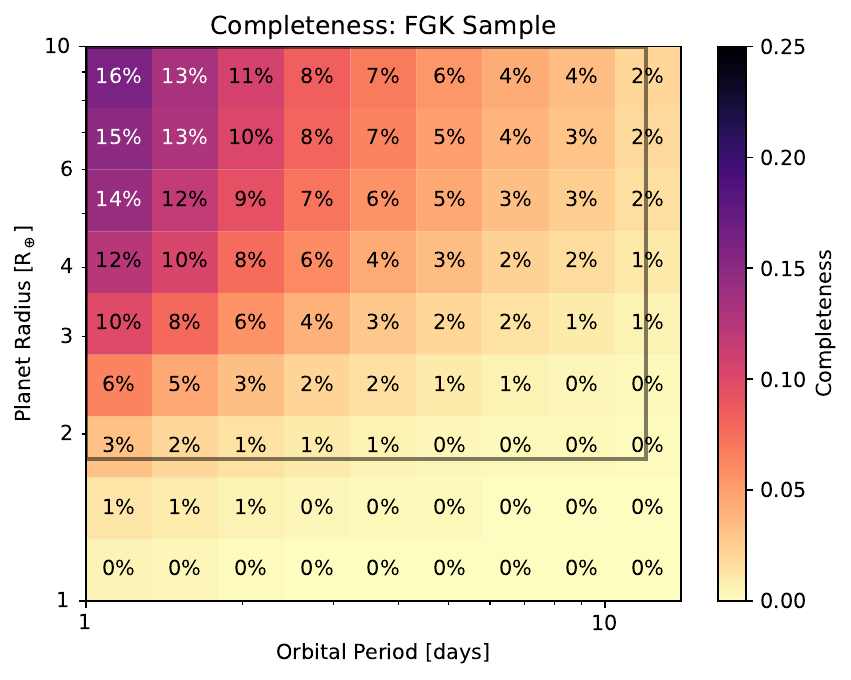}
\endminipage
\caption{\textbf{Top:} Overall detection efficiency of \pterodactyls in \rprs space for the full sample. \textbf{Center:} Detection efficiency of the FGK stars in our sample. \textbf{Bottom:} Completeness of the FGK sample which combines the detection efficiency with the geometric transit probability. The black box denotes the bin over which the intrinsic occurrence rates were calculated i.e., sub-Neptunes and Neptunes (1.8--10\,\Rearth) within 12\,days.}
\label{fig:det_eff}
\end{figure}

The detection efficiency of our survey was calculated using injection-recovery tests for planets with radii of 1--10\,\Rearth and orbital periods of 1--14\,days. For each star in our overall sample, we implemented a 3×4 injection-recovery (log-spaced) grid to balance computational efficiency with robust sampling of the parameter space. We interpolated results to a finer grid, achieving a higher-resolution detectability map for small variations in planetary properties. However, to better sample the detection efficiency for FGK stars, we performed injections on a finer 9×9 grid (81 injections per star), randomly drawing the P/R of each injection from uniform distributions within each grid cell. We also allowed the impact parameter b to vary between 0 and 0.9 for each injection. Because b is derived from the inclination, thus requiring a measure of stellar radius which we only derived for FGK stars, we assumed b=0 (central transit) while measuring the detection efficiency of the overall sample for which we do not calculate occurrence rates. For each injection, we randomly drew T0 from a uniform distribution between 0 and P, ensuring that all possible phases were sampled. This approach allowed us to explore the systemic noise and observing window properties of the ensemble of light curves. We injected these transit signals into each light curve in our sample and scaled the injection with the flux contribution of each star as calculated by \triceratops. We then passed these injections through \pterodactyls and calculated the fraction that were recovered as planets to produce a detection efficiency map per cluster as well as an average map for the entire sample. The average detection efficiency map for our entire sample of 5457 stars, and for the 1374 FGK stars in our sample is presented in the top and middle panels, respectively of Figure~\ref{fig:det_eff}. We find that the overall detection efficiency does not exceed 50\% for any bin below 0.1\,\rprs (akin to a Jupiter-sized planet orbiting a Sun-like star) or 10\,\Rearth, likely due to the effects of flux contamination \citep{fernandes2022}. In order to gauge how occurrence rates might change with age, we also calculate average detection efficiency maps for two age bins (10--100\,Myr and 100--1000\,Myr) using injection-recovery results for the stars of corresponding ages. These maps should also account for potential differences in observational biases between the age bins and clusters therein, such as variations in magnitude between young- and intermediate-age stars (Figure~\ref{fig:sample_comp}).

To move from detection efficiency to survey completeness, we also need to calculate the likelihood that a given planet will transit, given by the geometric transit probability:
\begin{equation}
f_\text{geo} = \frac{R_\star}{a}
\end{equation}
Here, $R_\star$ is the stellar radius, and $a$ is the planet semi-major axis which, through Kepler's third law, also incorporates a dependence on stellar mass. Because $f_\text{geo}$ depends on stellar properties, we calculate it per star in order to capture the variation in stellar properties over the range of our sample (e.g., Figure~\ref{fig:sample_comp}), and then multiply each with the sample's average detection efficiency map. This gives us a set of per-star completeness maps, which we then combine in three ways: one average map for the full sample (presented in Figure~\ref{fig:det_eff}) and two age-specific average maps for the 10-100\,Myr and 100-1000\,Myr age bins.

Using these completeness measurements, we can then provide estimates for the intrinsic planet occurrence rate ($\eta$). There are multiple ways to measure planet occurrence rates, and each technique carries both advantages and disadvantages. The inverse detection efficiency method provides direct measurements of planet occurrence rates in regimes with planet detections, but struggles to outline the distribution of occurrence rates with planet properties when samples are sparse. Meanwhile, occurrence rate forward modeling allows us to parametrically describe how occurrence rates are distributed, but requires imposing a specific functional form. We opt to use both techniques in the following subsections, so that we might better characterize -- both natively and parametrically -- the occurrence rates of our young planet sample.

\subsection{Inverse Detection Efficiency Method}\label{subsec:idem}

We first employed the inverse detection efficiency method, where occurrence rates $\eta$ are calculated in bins of orbital period and planet radius using:
\begin{equation}\label{eqn:IDEM}
    \eta_\mathrm{bin} = \frac{1}{n_*} \sum^{n_\mathrm{p}}_{j} \frac{1}{\mathrm{comp}_j},
\end{equation}
where $comp_\text{j}$ represents the survey completeness evaluated at the properties of the $j^\mathrm{th}$ planet, $n_p$ denotes the number of detected planets in the bin, and $n_\star$ stands for the number of surveyed stars. The uncertainty on the occurrence rate was computed from the square root of the number of detected planets in the bin, assuming Poissonian statistics.

In our analysis, we focused on planets with radii between 1.8--10\,\Rearth within 12\,days (roughly half a TESS sector) to examine the primordial population of sub-Neptunes and Neptunes. With an average detection efficiency of 40\%, we derived an occurrence rate of 54.0$\pm$16.3\% for our young cluster sample. 

For the same orbital period and planet radius bins, we also calculated an occurrence rate from the \kepler sample of Gyr-old FGK (Sun-like) stars. As in \citet{bergsten2022}, we employed the final \kepler{} data release \citep{thompson2018} supplemented with the \citet{Berger2020} catalog of \gaia-revised stellar properties \citep{gaiadr2}. We isolated a sample of 114,515 dwarf stars with masses between 0.56--1.63\,\Msun, based on FGK stellar mass range from \citet{pecautmamajek2013} and the \citet{Huber2016} criterion where a star is considered a dwarf if :
\begin{equation}
    \log{g} > \frac{1}{4.671}\arctan{\left(\frac{T_\mathrm{eff}-6300}{-67.172}\right)}+3.876.
\end{equation}
Among these stars are 554 confirmed and candidate planets  (as reported in \citealp{thompson2018}) with orbital periods between 1--12\,days and \gaia-updated planet radii between 1.8--10\,\Rearth.

Using this population of stars and planets, alongside a set of per-star completeness maps from \citet{bergsten2022}, we computed an occurrence rate value of 7.79$\pm$0.33\%, appreciably lower than our derived value for young clusters. If the measured decrease in planet occurrence is due to larger planets shrinking to smaller radii below our 1.8\,\Rearth cut-off, then it is possible that a \kepler occurrence rate which includes these smaller planets may be more similar to our young planet sample. To investigate this, we reevaluated occurrence rates from the \kepler sample while including planets down to 0.3\,\Rearth. From the inverse detection efficiency method, we find an occurrence of $39.5 \pm 1.0\%$ while including \kepler planets down to 0.3\,\Rearth. Thus, the derived young star planet occurrence rate  (54.0$\pm$16.3\%) is within 1$\sigma$ with that of \kepler assuming most planets lose their atmospheres over time.

\begin{figure*}[!htpb]
\minipage{\linewidth}
  \includegraphics[width=\linewidth]{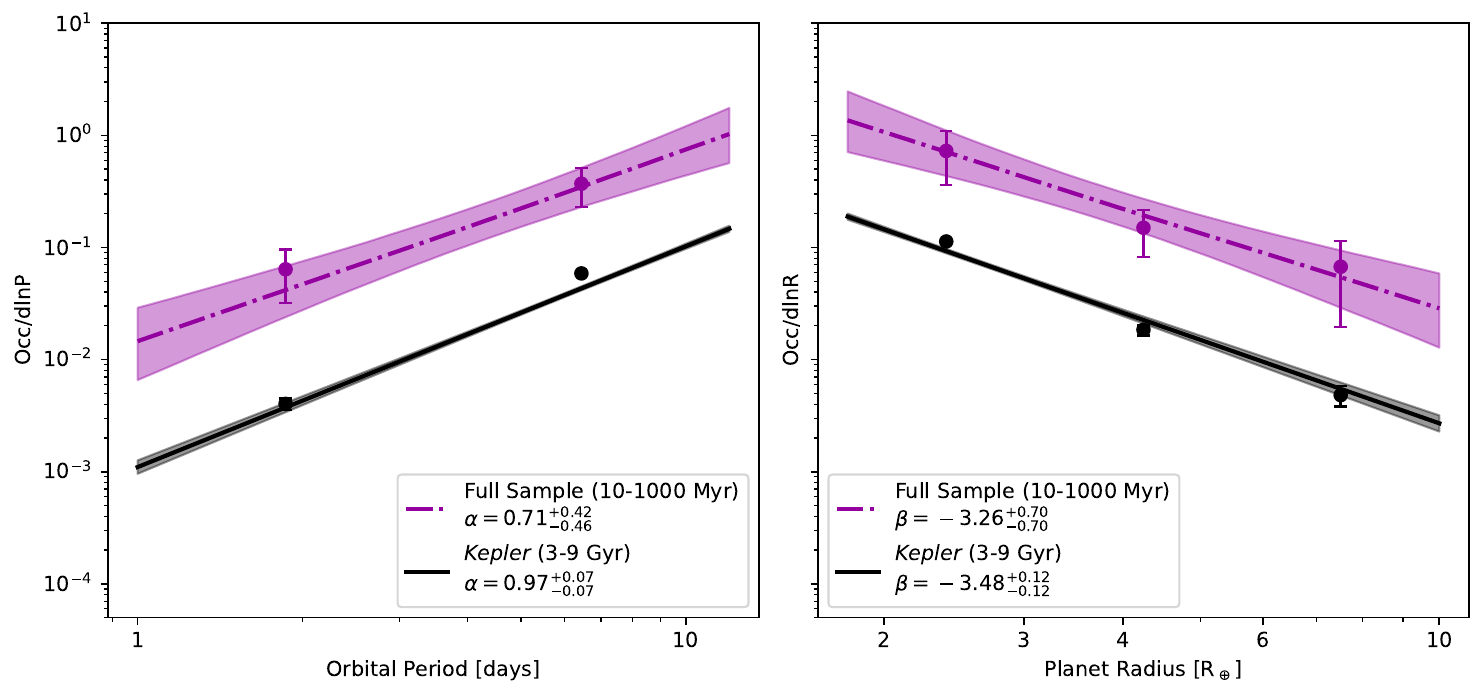}
\endminipage\hfill
\minipage{\linewidth}
  \includegraphics[width=\linewidth]{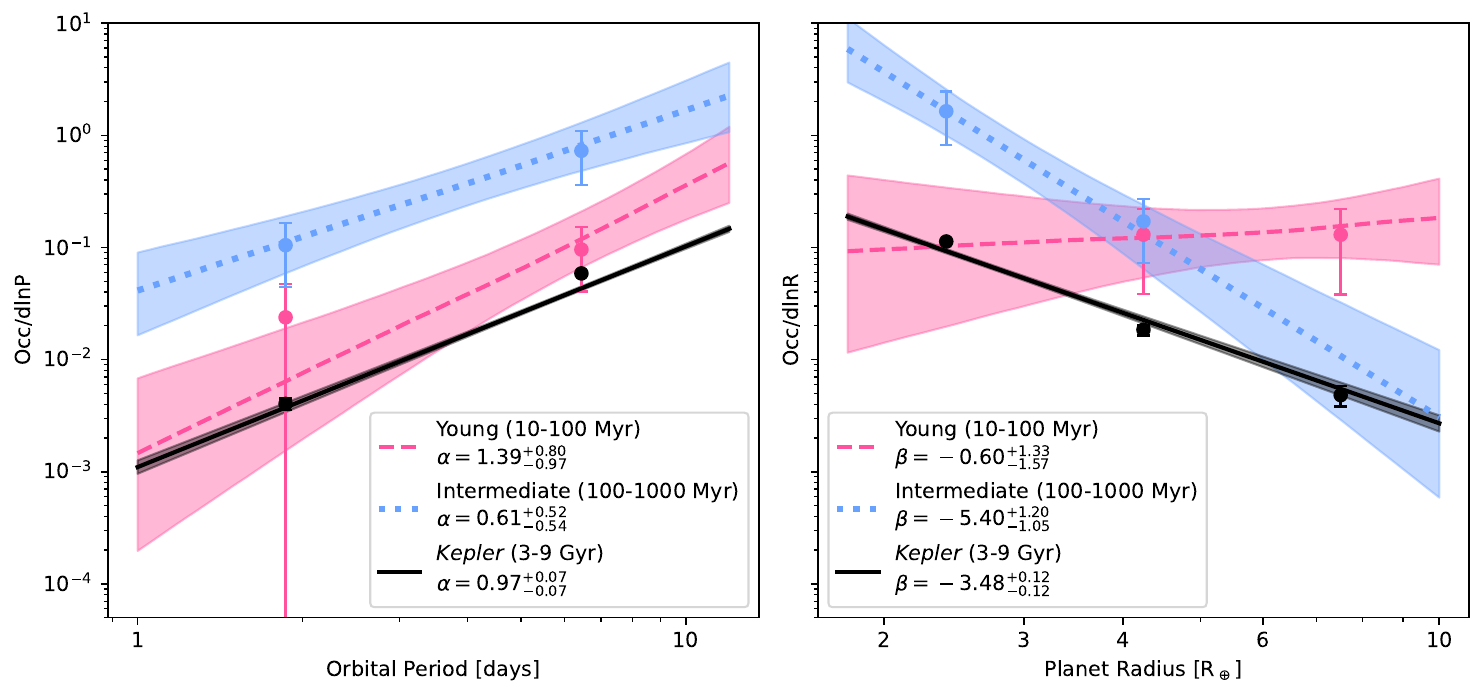}
\endminipage
\caption{Comparison of the \kepler's inverse detection efficiency vales (circles), our best fit occurrence models (lines) in orbital period and radius space, respectively, with \textbf{Top:} our entire sample (10--1000\,Myr), and \textbf{Bottom:} split into our young (10--100\,Myr) and intermediate (100--1000\,Myr).}
\label{fig:power_law_fit}
\end{figure*}

To explore how the occurrence of our sample might further depend on planet properties, we split our sample into two bins of equal log-width in orbital period and applied the inverse detection efficiency method; separately, we also repeated this using three bins of planet radius. The resulting values are plotted in Figure~\ref{fig:power_law_fit}, where we note a tendency for occurrence rates to increase with orbital period and decrease with planet radius. These trends have also been demonstrated in the \kepler Gyr-old population (see e.g., \citealp{petigura2018, bergsten2022}), as indicated in the \kepler values also plotted in Figure~\ref{fig:power_law_fit}.

Furthermore, we split our sample into two age bins with equal numbers of planets: 10--100\,Myr and 100--1000\,Myr. While the uncertainties on these points are necessarily larger than those of the full young sample, we note a $1\sigma$ distinct offset between values from the inverse detection efficiency method. To better understand and quantify the differences in these trends between our young sample and \kepler, we employ occurrence rate forward modeling to profile these trends over a more continuous distribution of periods and radii.

% \begin{figure}[!htb]
%     \centering
%     \includegraphics[width=\linewidth]{comp_per_FGK.pdf}
%     \caption{Caption}
% \label{fig:comp_per}
% \end{figure}

\subsection{Occurrence Rate Modeling}\label{subsec:modeling}

Separately from the previous subsection, we also fit parametric models to our young planet sample to better understand the distribution of planet occurrence rates. We adopt the approach of \citet{Youdin2011} and \citet{burke2015} to model the two-dimensional occurrence distribution, following the methodology of \citet{bergsten2022}.

Our model includes three free parameters: $F_0$ represents the average number of planets per star, while the parameters $\alpha$ and $\beta$ define the slope of a power law in orbital period and planet radius, respectively. Similar models will often use a broken power law in orbital period, although this break typically occurs somewhere around $\sim$10--12 days for FGK stars (see e.g., \citealp{mulders2018exoplanet, bergsten2022}). As our sample is limited to planets within 12\,days, a single power law is sufficient for our study. Our model thus takes the form:
\begin{equation}\label{eqn:pldf}
    \frac{\mathrm{d}^2 f}{\mathrm{d} P \mathrm{d} R} = F_\mathrm{0} C_\mathrm{n} P^\alpha R^\beta
\end{equation}
where $C_\mathrm{n}$ is the normalization constant from \citet{burke2015}. This function requires integrating the sum of each star's completeness (as described in Section~\ref{subsec:completeness}) over a grid of orbital periods and planet radii. Model parameters were optimized through a Markov-chain Monte Carlo process (\texttt{emcee}, \citealt{foreman2013}) to minimize the corresponding likelihood function (Eqn.~9 in \citealp{burke2015}). We used 64 walkers and ran for 20,000 steps (at least 50 times the autocorrelation time $\tau$), with the first 1,000 (at least 2$\tau$) discarded for burn-in.

To gauge if any model parameters changed with age, we fit separate models to three age bins: 10--100\,Myr, 100--1000\,Myr, and the full young sample. Additionally, to enable better comparison with the \kepler FGK sample, we include a model fit as described above, but using \kepler data and summed per-star completeness maps as in \citet{bergsten2022}.

\section{Results and Discussion}\label{sec:results}
\begin{figure*}[!htb]
    \centering
    \includegraphics[width=\linewidth]{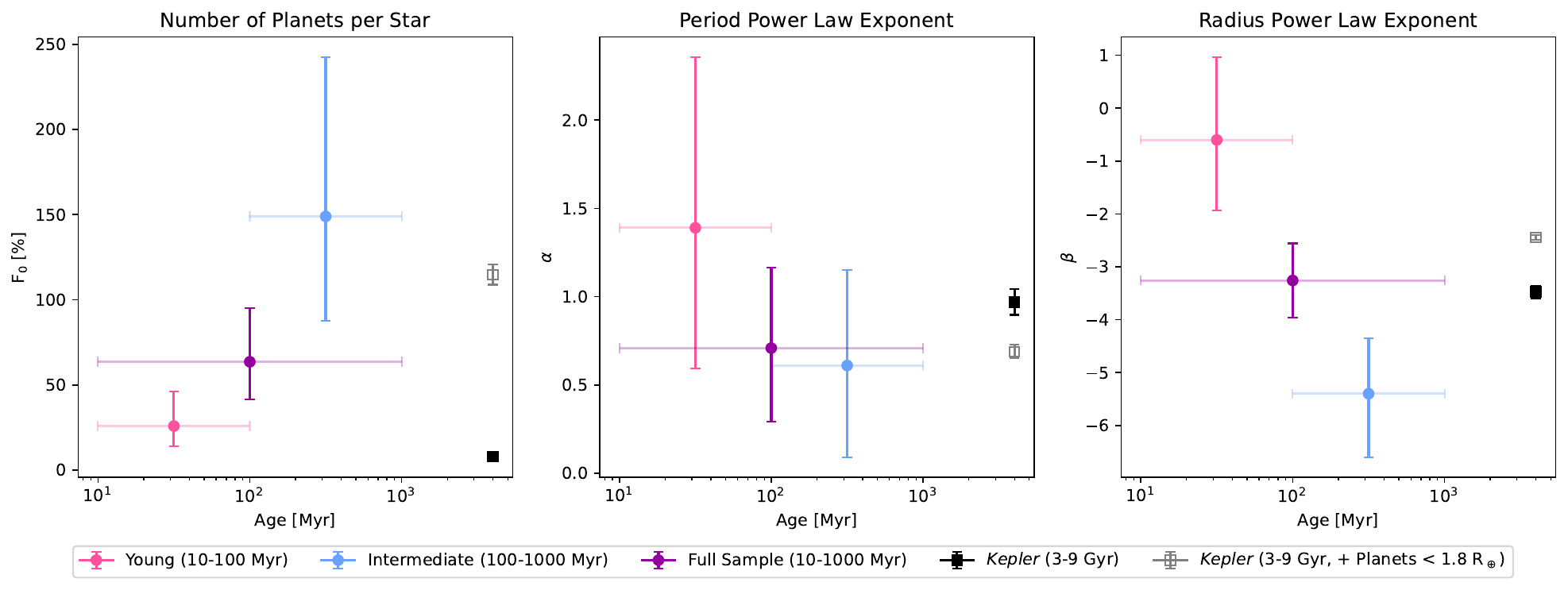}
    \caption{Comparison of the number of planets per star (F0), Period ($\alpha$), and Radius ($\beta$) exponents across the different age bins.}
\label{fig:power_law_comp}
\end{figure*}

In Figure~\ref{fig:power_law_fit}, we plot occurrence rates and uncertainties from the inverse detection efficiency method alongside our best-fit occurrence models. The model free parameters are plotted in Figure~\ref{fig:power_law_comp} and listed in Table~\ref{tab:fit_params}, with the median and lower/upper $1\sigma$ uncertainty values taken from the \nth{50} and \nth{16}/\nth{84} percentiles, respectively.

\begin{table*}[!htb]
    \centering
    \begin{tabular}{c|c|c|c}
         Sample (FGK) &  $F_0 [\%]$&  $\alpha$& $\beta$\\
         \hline
         10--100\,Myr&  ${25.89}_{-11.67}^{+20.18}$&  ${1.39}_{-0.80}^{+0.97}$& ${-0.60}_{-1.33}^{+1.57}$\\
         100--1000\,Myr&  ${148.93}_{-61.10}^{+93.60}$&  ${0.61}_{-0.52}^{+0.54}$& ${-5.40}_{-1.20}^{+1.05}$\\
         10--1000\,Myr&  ${63.62}_{-22.16}^{+31.58}$&  ${0.71}_{-0.42}^{+0.46}$& ${-3.26}_{-0.70}^{+0.70}$\\
         \kepler (Gyr-old) &  ${7.98}_{-0.35}^{+0.37}$&  ${0.97}_{-0.07}^{+0.07}$& ${-3.48}_{-0.12}^{+0.12}$\\
 \kepler (Gyr-old, + Planets $< 1.8$\,\Rearth)& ${114.70}_{-5.83}^{+6.00}$& ${0.69}_{-0.04}^{+0.04}$&${-2.44}_{-0.04}^{+0.04}$\\
    \end{tabular}
    \caption{Best-fit and $1\sigma$ model parameters for the two age bins of our sample, the full young sample, and the \kepler Gyr-old sample.}
    \label{tab:fit_params}
\end{table*}

Our models find that the occurrence rate of young planets is higher than the Gyr-old \kepler{} population, also seen with the inverse detection efficiency method as described in Section~\ref{subsec:idem}. From Table~\ref{tab:fit_params}, the average number of planets per star for our full young sample is $F_0 = {63.62}_{-22.16}^{+31.58}\%$, which is distinct from the \kepler{} value of ${7.98}_{-0.35}^{+0.37}\%$ at the $2.5\sigma$ level. Additionally, we measure $F_0 = {25.89}_{-11.67}^{+20.18}\%$ for our younger 10--100\,Myr age bin consistent with \citet{Vach2024}, and \textbf{$F_0 = {148.93}_{-61.10}^{+93.60}\%$} for our intermediate 100--1000\,Myr age bin consistent with \citet{Christiansen2023}. We note that the intermediate bin's value is larger and more uncertain than that of the younger, such that that our two age bins are distinct at the $1.9\sigma$ level (see e.g., Figure~\ref{fig:power_law_comp}). We present this difference between the age bins with caution, as it could be an artifact of small number statistics, or imposing a functional form that artificially inflates occurrence in the older bin (e.g., fitting a power law in radius space out to 10\,\Rearth despite no planets exceeding 4\,\Rearth in that bin).

Regarding the shape of the best-fit occurrence distributions, we measure a significant change in the slope of the occurrence-radius distribution with age, but not in the occurrence-period distribution. All of our best-fit occurrence-period slopes, for both the young stars and the \kepler{} sample(s), are consistent at $1\sigma$. Meanwhile, we find tentative ($2.2\sigma$) evidence that the occurrence-radius distribution may steepen with age, from $\beta = {-0.60}_{-1.33}^{+1.57}$ in our youngest age bin to ${-3.48}_{-0.12}^{+0.12}$ with \kepler{}. The latter is similar to the $\beta = {-3.26}_{-0.70}^{+0.70}$ value measured for our full young sample, and we neglect discussing the intermediate age bin which only has coverage of smaller ($<4$\,\Rearth) planets. 

\subsection{Interpretations and Implications}
\begin{figure*}[!htb]
    \centering
    \includegraphics[width=\linewidth]{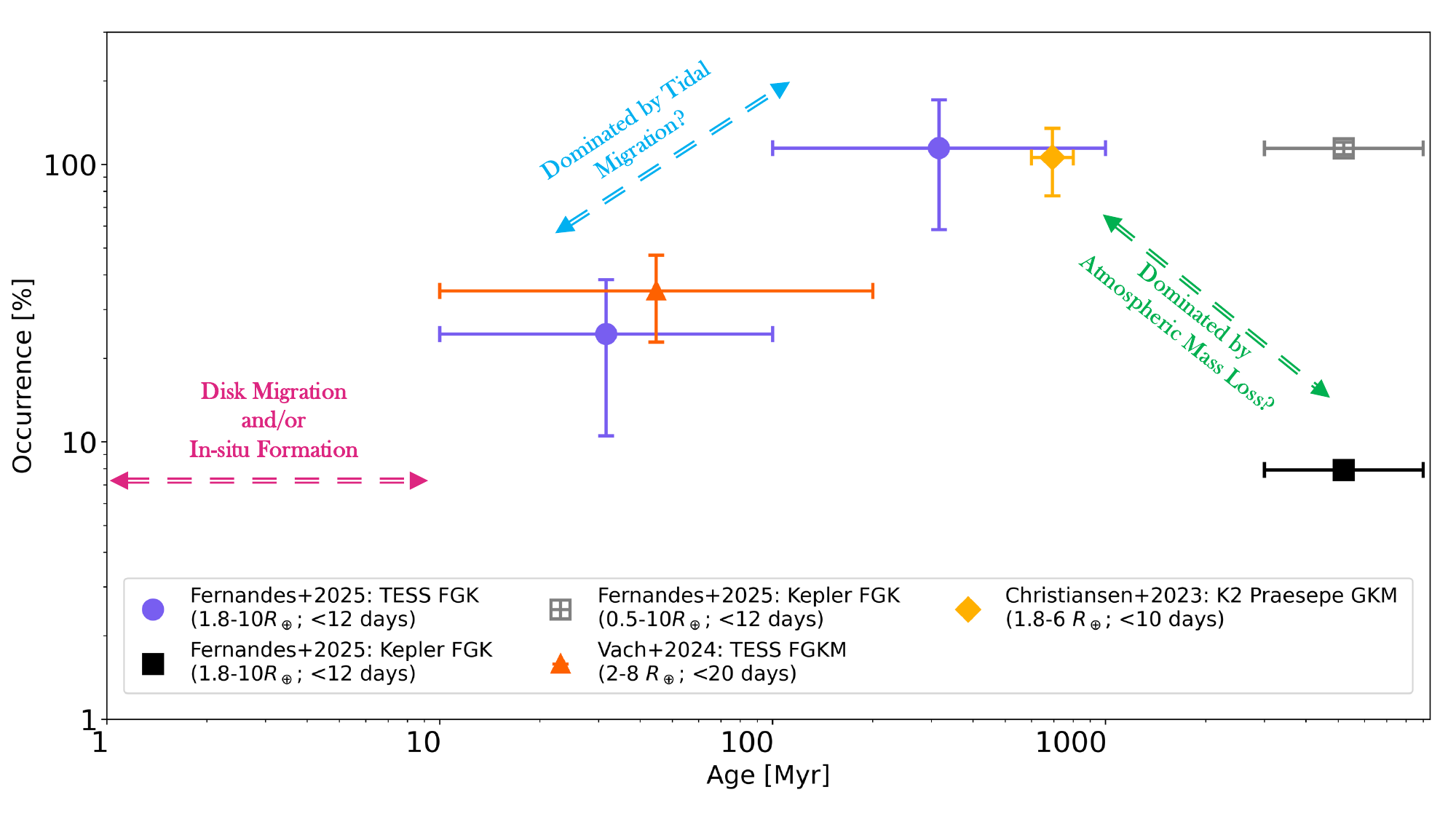}
    \caption{Comparison of occurrence rates for sub-Neptunes and Neptunes from different surveys, with the various underlying mechanisms and their timelines depicted with dashed arrows.}
\label{fig:occ_comp}
\end{figure*}
Here we discuss potential explanations for the results described above and their implications for planet formation/evolution. We precede this discussion with a note that our coverage of stellar age begins at 10\,Myr, offering only a partial view of planet formation and evolution. 

Evidence from older populations shows that the occurrence of sub-Neptunes ($<$4\,\Rearth) increases with orbital period up to $\sim$10\,days, beyond which it flattens in log-period space, forming a plateau across all stellar types (F, G, K, and M) in Gyr-old dwarfs \citep{mulders2015planet}. This turnover likely reflects the gas disk’s truncation by the host star’s magnetosphere at the corotation radius (see also \citealt{lee17}), suggesting that planets often migrate to the gas disk truncation radius early on. These findings suggest that early disk-driven migration, either of fully formed planets or their building blocks, plays a key role in shaping the inner architecture of planetary systems and that planets with periods $<$10 days have been most influenced by stellar tides over time. Our survey is sensitive to young ($<$1\,Gyr) sub-Neptunes primarily inside 10\,days, and the finding of increased occurrence between the 10--100\,Myr and 100--1000\,Myr age bins aligns with the long timescales required for tides to bring planets within the disk truncation radius (e.g., \citealt{Jackson2009}). The presence of planets with periods $<$10 days in our youngest age bin suggests that migration to $-$ or formation at $-$ such short periods must occur within the first 10--100\,Myr. Given the lack of substantial material within the truncation radius, in-situ formation of planets at such short periods is unlikely but cannot be entirely ruled out.

Next, the observed increase in occurrence between our younger and intermediate age bins, from $F_0 = {25.89}_{-11.67}^{+20.18}\%$ to ${148.93}_{-61.10}^{+93.60}\%$, could be indicative of further migration. If the majority of large, short-period planets form or arrive at their final location during the gas disk stage, we would expect a decline in occurrence rates with age driven by atmospheric cooling and escape. The observed rise in occurrence rates here would thus suggest a later arrival of planets between 100\,Myr and 1\,Gyr, possibly through tidal migration. If a planet is disturbed onto an elliptical orbit through processes like planet-planet scattering (e.g., \citealt{rasi96,weid96}), secular cycles (e.g., \citealt{wu03}), or secular chaos (e.g., \citealt{petr19}), it can migrate via tides (e.g., \citealt{hut81,eggleton1998}) raised on the planet (e.g., \citealt{socr12b}). Regardless of orbital eccentricity, a planet can also migrate via tides raised on the star. Stellar tides could further reduce a planet's period beyond the truncation radius to explain the population of planets inside 10\,days \citep{lee17} which our survey is mostly sensitive to. The processes disturbing planets into elliptical orbits can operate over a range of timescales. Moreover, the planet’s initial position and eccentricity—as well as uncertain quantities like the tidal quality factor—set the tidal migration timescale, which can span many orders of magnitude. Thus, we expect new contributions from tidal migration in every order of magnitude of age (e.g., as argued by \citealt{safs20}).

One caveat is that migration would also likely cause the occurrence-period slope to become more shallow with time, as planets become less sharply-peaked near the truncation radius and more evenly distributed across shorter periods. Unfortunately, we cannot provide complement the observed trend in $F_0$ with evidence for the evolving occurrence-period slope due to large uncertainties driven by small sample sizes. While a steeper slope at young ages is plausible (Figure~\ref{fig:power_law_comp}) within our uncertainties, a slope that does not change with age is also possible, so constraining the true behavior of this parameter ultimately requires a larger dataset. If the slope remains unchanged with age, it may suggest that planets within 1–12 days either do not migrate or experience a migratory process that acts uniformly in log-period space.

Moving from our young sample to the Gyr-old \kepler{} population, the moderately significant decrease in occurrence between the intermediate age bin and the \kepler{} sample (from $F_0 = {148.93}_{-61.10}^{+93.60}$ to ${7.98}_{-0.35}^{+0.37}$) could be indicative of planetary atmospheric mass loss. Atmospheric thermal evolution and mass loss play important roles in controlling the size distribution of planets hosting H/He-dominated atmospheres. Many studies have shown that planetary atmospheres can significantly contract with time, as a direct result of thermal cooling and stellar-driven escape \citep[e.g.][]{owen2013kepler,lopez2013role}. Of note, core-powered mass loss relies on a bolometrically heated outflow \citep{gupta2019sculpting}, whereas photoevaporative mass loss relies on XUV-driven outflow \citep{Rogers2021a}. 

While the complex physical interplay between core-powered and photoevaporative mass loss is an area of active research \citep[e.g.][]{Owen2024,Rogers2024}, one would expect the planet demographics to change with time as a result of these processes. Specifically, there should be a higher abundance of large planets that will eventually contract to become sub-Neptunes, or super-Earths if their atmospheres are completely stripped. Our results agree with this general trend, demonstrated in Figure \ref{fig:power_law_fit} and the evolution of free parameters for the occurrence $F_0$ and occurrence-radius slope $\beta$ (Figure~\ref{fig:power_law_comp}, Table~\ref{tab:fit_params}). Comparing the same 1.8--10\,\Rearth range for our intermediate age bin and \kepler, we see a higher planet occurrence around young stars with a flatter occurrence-radius distribution (both true at $2.5\sigma$), indicating an increased fraction of puffier planets. When planets smaller than 1.8\,\Rearth (i.e., the super-Earths) are accounted for, the occurrence of Gyr-old planets becomes comparable to that of intermediate-age planets ($F_0 = {114.70}_{-5.83}^{+6.00}$ and ${148.93}_{-61.10}^{+93.60}$, respectively), consistent with the idea of planets shrinking below our $1.8$\,\Rearth cutoff with age due to atmospheric mass loss processes.

Given that radius contraction due to thermal cooling is not expected to significantly shrink planets and most of it occurs before 1\,Gyr (e.g., \citealt{mordasini2012}), this suggests that gas-rich planets are losing their atmospheres to become Super-Earths on Gyr-timescales, consistent with previous observational studies \citep{berger2020b,trevor2021a}. This poses an apparent challenge to photoevaporation models because the bulk of the atmospheric loss due to photoevaporation is often predicted to take place during the first 100\,Myr \citep{Rogers2021a} when a significant fraction of the planets may still be migrating to their final location. This view of the problem, however, may be overly simplistic because both photoevaporation and core-powered mass loss continue at varying levels on Gyr-timescales. Coupled models of planet formation and envelope loss due to different mechanisms will help to shed light on planet evolution based on our results. We note that, since our completeness is too low for young, small planets below 1.8\,\Rearth, we are unable to probe any increase in occurrence of super-Earths, as would be predicted by mass loss models. However, we note that this process is critical to take into account to properly estimate the occurrence of Earth-size planets in the habitable zone \citep{Pascucci2019, bergsten2022}.

Tidal disruption may also play a role, destroying either the entire planet or its atmosphere (e.g., \citealt{fabe05,guil11}). Planets with orbital periods shorter than the stellar spin period transfer angular momentum to spin up their stars, so tidal disruption is the ultimate fate of migration driven by tides raised on the star. Tidal disruption is also the ultimate fate of tidal migration driven by tides raised on the planet if the planet’s periapse comes within the Roche limit. As an end state of tidal migration, tidal disruption occurs over many orders of magnitude in timescales. It is not obvious that the balance between arriving and disrupted planets would raise the occurrence rate from 100\,Myr to 1\,Gyr but substantially drop it at 1--10\,Gyr. Therefore, it seems unlikely that tidal disruption is the predominant cause of the drop.

Another explanation for a change in the occurrence rate of planets with time is not that the planets themselves are evolving, but that planetary systems themselves form differently at different Galactic epochs. \cite{Christiansen2023} and references therein discuss a variety of Galactic environmental factors that could impact planet formation that have changed with time. Stars born at earlier times can be significantly more strongly externally irradiated by higher Galactic star formation rates, which could reduce both protoplanetary disk lifetimes and the materials remaining for planet formation. Stars born at earlier times are also typically born in more highly dynamically clustered environments, which could lead to frequent interactions between nearby stars that could destabilize the planetary systems that did form. Unlike the mass-loss mechanisms described above, there are currently few published predictions of the extent to which these processes may impact planet formation to test, however at this stage these effects cannot be ruled out as the cause of some of the trends seen here. 
%%%%%%%%%%%%%%%%%%%%%%%%%%%%%%%%%%%%%%%%%%%%%%%%%%%%%%%%%%%%%%%%%%%%%

\subsection{Hot Neptune Desert}
% \begin{figure}[!htb]
%     \centering
%     \includegraphics[width=\linewidth]{HotNeps.pdf}
%     \caption{caption}
% \label{fig:hotneps}
% \end{figure}
We also calculated occurrence rates in the hot Neptune desert, here defined as planets with radii between 3--10\,\Rearth and orbital periods between 1--4\,days \citep{Beauge2013}. The hot Neptune desert hosts some of the most highly irradiated exoplanets. As such, we expect atmospheric escape to be a dominant driver in the radius evolution of planets within the desert. Using the inverse detection efficiency method, we find an occurrence rate of $6.28\pm3.14\%$ for young planets in our sample. Further dividing our sample, we find $2.96\pm2.96\%$ for the occurrence rate of planets between 10--100\,Myr (note that $n=1$ here), and the occurrence rate of $9.76\pm5.64\%$ for between 100--1000\,Myr. The occurrence rate of young planets is therefore higher than the occurrence rate of $0.19\pm0.04\%$ from \kepler (1.9$\sigma$), which supports the notion that atmospheric escape is dominant for such planets, causing them to contract below $\sim$1.8\,\Rearth over time. This result, however, is only weakly significant. Larger sample sizes would reduce the uncertainties, providing a more clear picture of how this region of planet parameter space evolves with age. In addition, we note that our survey is not sensitive to planets below 1.8\,\Rearth, so a more sensitive characterization of smaller planets in young clusters may provide more insights.

% Beyond atmosphere loss and thermal cooling, other mechanisms could also be at play, that cause a change in the planet occurrence, especially in the 10-1000 Myr age range. Examples include post-disc migration, post-disc formation and varying stellar and galactic environments for different stellar clusters. We highlight again, however, that caution should be taken when drawing conclusions from small-number statistics.

% \begin{itemize}

%     \item Does the completeness estimates take into account that planets around younger stars are harder to detect because they might be active? If not, then if this could actually be accounted for -- it would increase the occurrence of young planets, especially the small ones! No?

%     \item would be interesting to see how much the Kepler occurrence increases if 1-10 REarth planets are accounted for. If the difference between 1.8-10 and 1-10 is comparable to the difference between the black and green data points on Figure 5b -- that then demonstrates that these stellar clusters might not be that different from the KEpler planets/stars. If not, they would be or some planet engulfment is taking place as Bekki suggests below.
% \end{itemize}

\section{Summary}\label{sec:summary}
In this study, we expanded on our work in \cite{fernandes2022} and \cite{fernandes_khu_2023}, and calculate the occurrence rates of short-period sub-Neptunes and Neptunes in nearby ($<$200\,pc) young clusters and moving groups (10\,Myr--1\,Gyr) using TESS EM1 FFIs. These planets provide insight into a population much closer in time to the primordial planet population. By comparing the intrinsic occurrence rates of the young planets with that of the mature \kepler population, we can begin to piece together how planetary systems have evolved with time. In our analysis, we find that:
\begin{itemize}[leftmargin=*]
    \item For our entire sample (10\,Myr--1\,Gyr), the intrinsic occurrence rate for young, short-period sub-Neptunes and Neptunes (${63.62}_{-22.16}^{+31.58}$\%) is higher than that of older planets observed by \kepler (${7.98}_{-0.35}^{+0.37}$\%), for the same period and radius bins, confirming previous results \citep[e.g.,][]{Christiansen2023, Vach2024}.

    \item Binning our sample by age, the occurrence increases from young (10--100\,Myr) to intermediate (100--1000\,Myr) ages, from ${25.89}_{-11.67}^{+20.18}\%$ to ${148.93}_{-61.10}^{+93.60}\%$. This distinction is significant at the $1.9\sigma$ level and, if real, suggests a later arrival of planets between 100\,Myr and 1\,Gyr, possibly through tidal migration in which planets perturbed onto elliptical orbits can migrate via tides raised on the planet or the star. Tidal migration processes can operate over a range of timescales, contributing to occurrence rates in different age bins. If tidal migration increases rates from 100\,Myr to 1\,Gyr, it must be masked by other declines in the 1-10\,Gyr bin.

    \item The occurrence also decreases from $F_0 = {148.93}_{-61.10}^{+93.60}$ at intermediate ages to ${7.98}_{-0.35}^{+0.37}$ with \kepler{}. We also observe a steepening of the occurrence-radius distribution with age (from $\beta = {-0.60}_{-1.33}^{+1.57}$ in our youngest age bin to ${-3.48}_{-0.12}^{+0.12}$ with \kepler{}). These results indicate a population of planets undergoing atmospheric thermal evolution and mass loss, including core-powered and photoevaporative mass loss. These processes play important roles in controlling the size distribution of H/He-dominated planets, and cause planets to contract over time due to thermal cooling and stellar-driven escape. 

    \item The occurrence rate in the hot Neptune desert (3--10\,\Rearth, 1--4\,days), using inverse detection efficiency, is $6.28\pm3.14\%$ for young planets (10\,Myr--1\,Gyr). This is significantly higher than \kepler's $0.19\pm0.04\%$, suggesting that planets that initially formed in the hot Neptune Desert may shrink below 1.8\,\Rearth due to extreme atmospheric escape.

\end{itemize}
While we find evidence of post-disk migration and atmospheric mass loss sculpting the population of short-period planets over time, there is still much left to be done. For example, we are unable to determine whether the occurrence-period distribution changes with age due to large uncertainties driven by our small sample size. Overall, there is a need to detect more young planets. Currently, there are fewer than 40 young planets found in both clusters/moving groups and young field stars, compared to thousands around Gyr-old field stars with \kepler. Increasing the sample of young planets will enable a more statistically robust comparison of the intrinsic occurrence rates with time.

There is a need to simultaneously compute the occurrence of longer-period, young planets ($>$12\,days), and compare it with that of short-period young planets to fully understand the role of post-disk migration. To evaluate if atmospheric mass loss is a dominant driver in the radius evolution of gas-rich planets we need to detect planets smaller than 1.8\,\Rearth (super-Earths) and compare their occurrence with that of gas-rich planets over time. Given that the light curves of young stars are highly variable, making it challenging to detect small planets, we need more precise light curves, which upcoming missions like PLATO \citep{Rauer2024} can likely provide.

Furthermore, the atmospheres of young planets can help provide context on the conditions under which planets form. By studying the volatile composition of young planetary atmospheres, we can begin to distinguish between in-situ and ex-situ (plus migration) scenarios, placing constraints on the timescales of planet formation. The composition of these planetary atmospheres, paired with the activity levels of the host stars, can help us better understand the efficiency and timescales of atmospheric mass loss mechanisms. Through these investigations, we can significantly enhance our understanding of planetary formation, migration, and evolution processes, providing a clearer picture of the dynamic history of planetary systems.

%%%%%%%%%%%%%%%%%%%%%%%%%%%%%%%%%%%%%%%%%%%%%%%%%%%%%%%
%%%%%%%%%%%%%%%%%%% Software %%%%%%%%%%%%%%%%%%%%%%%
%%%%%%%%%%%%%%%%%%%%%%%%%%%%%%%%%%%%%%%%%%%%%%%%%%%%%%%
\software{\pterodactyls \citep{rachel_fernandes_2022_6667960}, \texttt{NumPy} \citep{numpy}, \texttt{SciPy} \citep{scipy}, \texttt{Matplotlib} \citep{pyplot}, \eleanor{} \citep{feinstein2019eleanor}, \wotan{} \citep{hippke2019wotan}, \texttt{transitleastsquares} \citep{hippke2019optimized}, \texttt{vetting} \citep{hedges2021vetting}, \triceratops{} \citep{giacalone2020vetting}, \texttt{EDI-Vetter Unplugged} \citep{zink2019edivetter}, \exotic \citep{zellem2020utilizing}, \epos \citep{Mulders2018}}, \texttt{emcee} \citep{foreman2013}, \texttt{corner} \citep{corner}

%%%%%%%%%%%%%%%%%%%%%%%%%%%%%%%%%%%%%%%%%%%%%%%%%%%%%%%
%%%%%%%%%%%%%%%%%%% ACKNOWLEDGMENTS %%%%%%%%%%%%%%%%%%%%%%%
%%%%%%%%%%%%%%%%%%%%%%%%%%%%%%%%%%%%%%%%%%%%%%%%%%%%%%%

\section{Acknowledgements}
R.B.F. would like to thank Eric Ford for their expertise, assistance and, invaluable insights throughout this work. I.P., G.B., and K. C. acknowledge support from the NASA Astrophysics Data Analysis Program under grant  No. 80NSSC20K0446. G.D.M. acknowledges support from FONDECYT project 11221206, from ANID --- Millennium Science Initiative --- ICN12\_009, and the ANID BASAL project FB210003. S.G. is supported by an NSF Astronomy and Astrophysics Postdoctoral Fellowship under award AST-2303922. A.G. is grateful to the Heising-Simons Foundation for the 51 Pegasi b Fellowship and to Princeton University for the Harry H. Hess Fellowship and the Future Faculty in Physical Sciences Fellowship. T.T.K acknowledges support by the NASA/XRP grant 80NSSC23K0260.

The Center for Exoplanets and Habitable Worlds and the Penn State Extraterrestrial Intelligence Center are supported by Penn State and its Eberly College of Science. Computations for this research were performed on the Pennsylvania State University’s Institute for Computational and Data Sciences’ Roar supercomputer. Part of this research was carried out in part at the Jet Propulsion Laboratory, California Institute of Technology, under a contract with the National Aeronautics and Space Administration (80NM0018D0004). This paper includes data collected by the TESS mission. Funding for the TESS mission is provided by the NASA's Science Mission Directorate. This material is based upon work supported by the National Aeronautics and Space Administration under Agreement No. 80NSSC21K0593 for the program “Alien Earths”. The results reported herein benefited from collaborations and/or information exchange within NASA’s Nexus for Exoplanet System Science (NExSS) research coordination network sponsored by NASA’s Science Mission Directorate. This research has made use of the NASA Exoplanet Archive, which is operated by the California Institute of Technology, under contract with the National Aeronautics and Space Administration under the Exoplanet Exploration Program.
\clearpage

%%%%%%%%%%%%%%%%%%%%%%%%%%%%%%%%%%%%%%%%%%%%%%%%%%%%%%%
%%%%%%%%%%%%%%%%%%% BIBLIOGRAPHY %%%%%%%%%%%%%%%%%%%%%%%
%%%%%%%%%%%%%%%%%%%%%%%%%%%%%%%%%%%%%%%%%%%%%%%%%%%%%%%
\bibliographystyle{apj}
\bibliography{main}

%%%%%%%%%%%%%%%%%%%%%%%%%%%%%%%%%%%%%%%%%%%%%%%%%%%%%%%
%%%%%%%%%%%%%%%%%%% APPENDIX %%%%%%%%%%%%%%%%%%%%%%%
%%%%%%%%%%%%%%%%%%%%%%%%%%%%%%%%%%%%%%%%%%%%%%%%%%%%%%%
\appendix
\section{Comparison of \eleanor vs SPOC} \label{app:lc_comp}
We compare the RMS of \texttt{eleanor}, \texttt{SPOC}, and \texttt{QLP} light curves for planet-hosting stars, showing that \texttt{SPOC} light curves generally exhibit the lowest noise levels, making them the most suitable for planet detection. Given their reduced RMS residual scatter relative to \texttt{eleanor}, we chose \texttt{TESS SPOC FFI PDCSAP} light curves to improve our sensitivity to small planets in these intrinsically variable data sets.

\begin{figure*}[!htb]
    \centering
    \includegraphics[width=\linewidth]{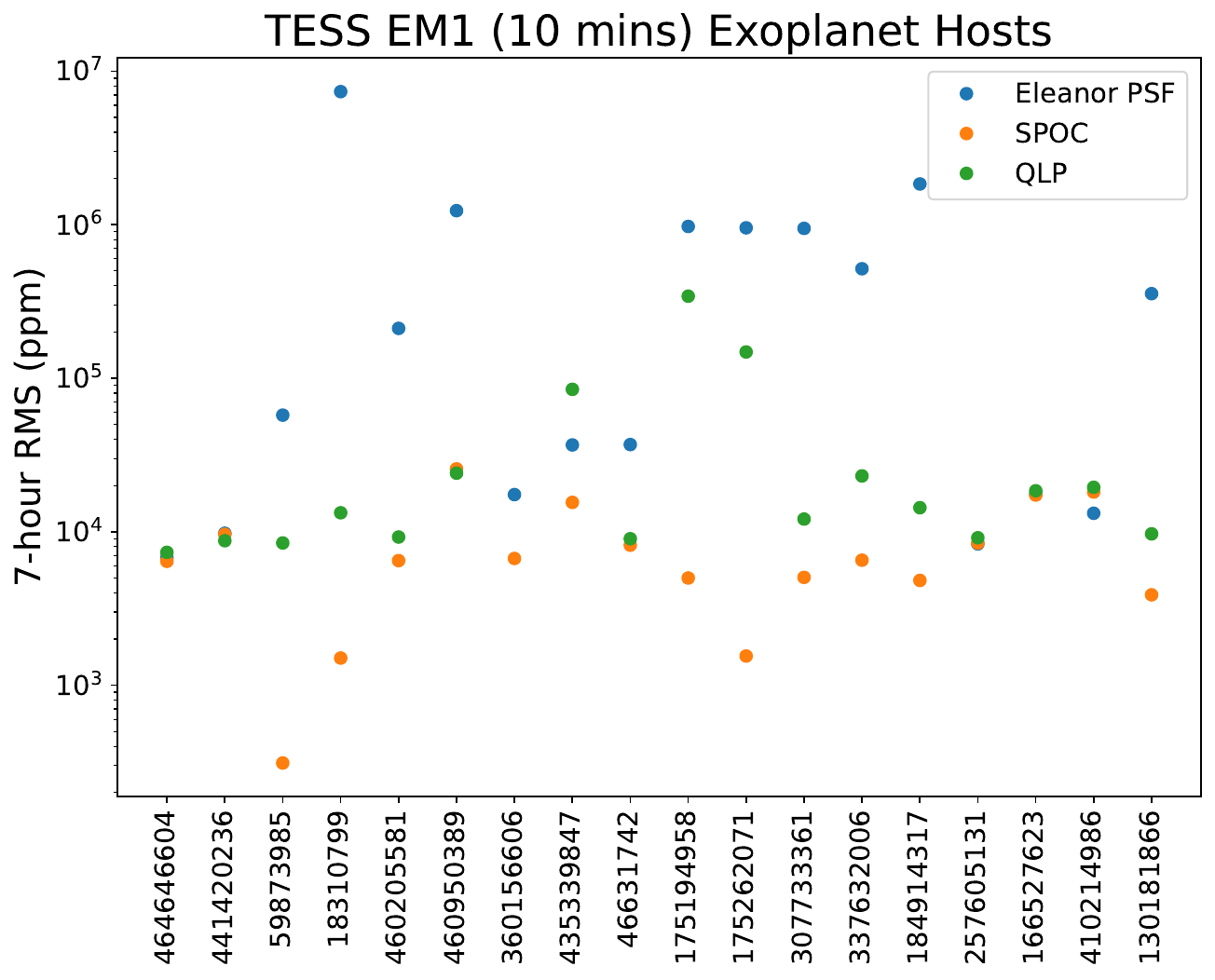}
    \caption{Comparison of RMS of \eleanor, SPOC and QLP light curves for the planet hosts depicting that, on average, SPOC light curves have the least amount of noise thereby making them ideal for planet searches.}
\label{fig:lc_comp}
\end{figure*}

\clearpage
\section{Skye Excess Metric (SEM)}\label{app:sem}
To address cadences with an unusually high number of transit-like detections, we applied a test similar to the “Skye'' metric for sectors 27-55, assessing transit-like signals at each cadence. Any cadences where the signal count exceeded 3$\sigma$ were masked before reprocessing the search to mitigate issues with the light curve.
\begin{figure*}[!htb]
    \centering
    \includegraphics[width=\linewidth]{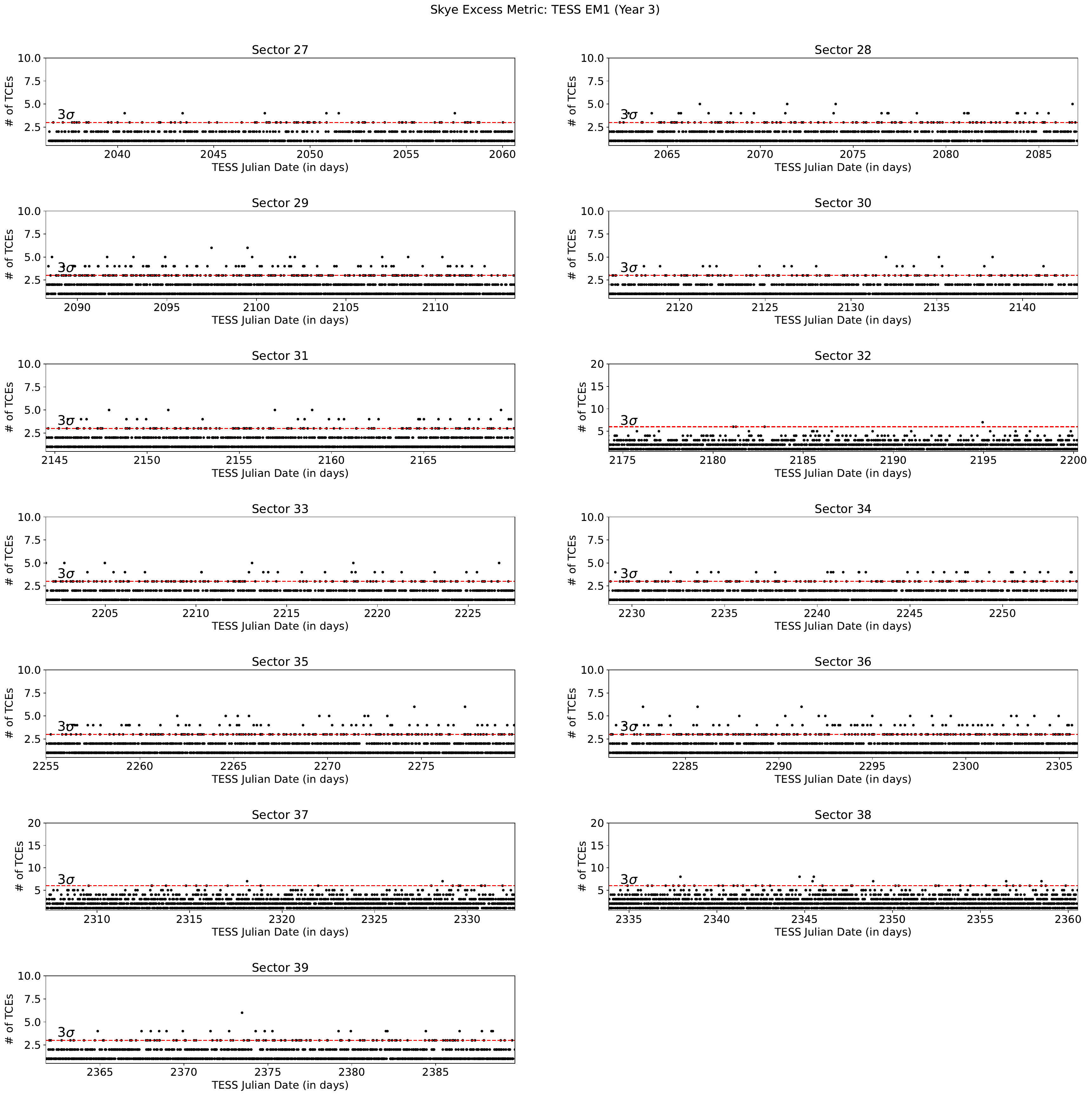}
    \caption{SEM for TESS EM1's Sectors 27-39}
\label{fig:sem_y3}
\end{figure*}

\begin{figure*}[!htb]
    \centering
    \includegraphics[width=\linewidth]{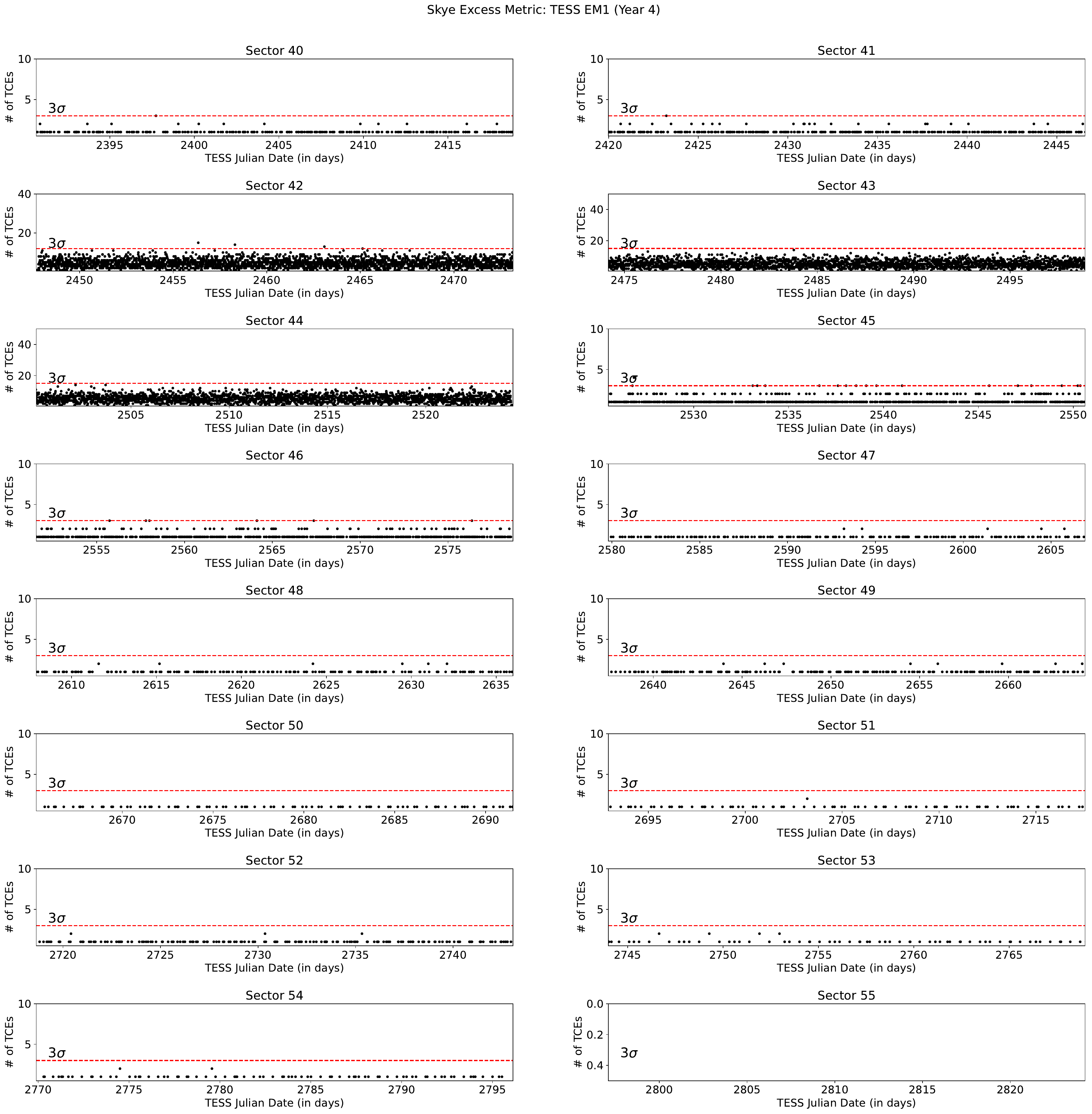}
    \caption{SEM for TESS EM1's Sectors 40-55}
\label{fig:sem_y4}
\end{figure*}

\end{document}